\documentclass[useAMS,usenatbib]{mn2e}
\usepackage{graphicx}
\usepackage{journal_names}

\usepackage{subfig}
\usepackage{amsmath}
\usepackage{amssymb}

\title[Novalikes CVs are Significant Radio Emitters]{Novalike Cataclysmic Variables are Significant Radio Emitters}
\author[Deanne L. Coppejans et al.]{Deanne L. Coppejans$^{1}$\thanks{Email: d.coppejans@astro.ru.nl}, Elmar G. K\"{o}rding$^{1}$, James C.A. Miller-Jones$^{2}$, \newauthor Michael P. Rupen$^{3}$, Christian Knigge$^{4}$, Gregory R. Sivakoff$^{5}$, Paul J. Groot$^1$\\
$^{1}$Department of Astrophysics/IMAPP, Radboud University, P.O. Box 9010, 6500 GL Nijmegen, The Netherlands\\
$^{2}$International Centre for Radio Astronomy Research, Curtin University, GPO Box U1987, Perth, WA 6845, Australia\\
$^{3}$National Research Council of Canada, Herzberg Astronomy and Astrophysics Programs,\\Dominion Radio Astrophysical Observatory, P.O. Box 248, Penticton, BC V2A 6J9, Canada\\
$^{4}$School of Physics and Astronomy, Southampton University, Highfield, Southampton SO17 1BJ, UK\\
$^{5}$Department of Physics, University of Alberta, CCIS 4-183, Edmonton, Alberta T6G 2E1, Canada}
\begin{document}

\date{}

\pagerange{\pageref{firstpage}--\pageref{lastpage}} \pubyear{2014}

\maketitle

\label{firstpage}

\begin{abstract}

Radio emission from non-magnetic cataclysmic variables (CVs, accreting white dwarfs) could allow detailed studies of outflows and possibly accretion flows in these nearby, numerous and non-relativistic compact accretors. Up to now, however, very few CVs have been detected in the radio.

We have conducted a VLA pilot survey of four close and optically-bright novalike CVs at 6 GHz, detecting three, and thereby doubling the number of radio detections of these systems. RW Sex, V603 Aql and the old nova TT Ari were detected in both of the epochs, while V1084 Her was not detected (to a $3\sigma$ upper-limit of 7.8 $\mu\rm{Jy}\,\rm{beam}^{-1}$). These observations clearly show that the sensitivity of previous surveys was typically too low to detect these objects and that non-magnetic CVs can indeed be significant radio emitters.

The three detected sources show a range of properties, including flaring and variability on both short ($\sim$200 s) and longer-term (days) time-scales, as well as circular polarization levels of up to 100\%. The spectral indices range from steep to inverted; TT Ari shows a spectral turnover at $\sim$6.5 GHz, while the spectral index of V603 Aql flattened from $\alpha=0.54\pm0.05$ to $0.16\pm0.08$ ($F_{\nu}\propto \nu^{\alpha}$) in the week between observations. This range of properties suggests that more than one emission process can be responsible for the radio emission in non-magnetic CVs. In this sample we find that individual systems are consistent with optically thick synchrotron emission, gyrosynchrotron emission or cyclotron maser emission.
 
\end{abstract}


\section{Introduction}

Radio emission is found, at least intermittently, from nearly all kinds of accreting objects. The most prominent radio emitters in the Universe are radio-loud Active Galactic Nuclei (AGN) - accreting supermassive black holes. AGN produce tightly collimated jets that are responsible for the majority of their radio emission. Accreting stellar mass black holes, the X-ray binaries (XRB), also show radio emission during particular stages of the outburst cycle \citep{Fender2001,Fender2004}. The same holds true for neutron star XRBs \citep{Migliari2006}.

Scaling relations connecting different classes of black holes have been found \citep{Merloni2003,Falcke2004}. This suggests that accretion and its associated phenomena can -- at least to a first order approximation -- be scaled from one source class to the other, and notably to accreting white dwarfs (WDs) \citep{Koerding2006,Koerding2007,Koerding2008b}. As WDs are nearby, numerous and non-relativistic they are ideal laboratories for accretion studies in compact objects (see e.g. \citealt{Martino2015}).

Cataclysmic variable stars (CVs) are the nearest examples of accreting compact objects. These binary star systems comprise a white dwarf that accretes matter from a red dwarf secondary star via Roche-lobe overflow (see \citealt{Warner1995}). CVs are broadly classified according to their magnetic field strength (B) into the magnetic systems, namely polars ($B\gtrsim10^7$G) and intermediate polars ($10^6\lesssim B\lesssim10^7\,\textrm{G}$), and non-magnetic systems ($B\lesssim10^6\,\textrm{G}$). The WDs in the non-magnetic systems accrete directly via an accretion disc onto the surface of the WD, but in the intermediate polars (IPs) the disc is truncated at the Alf\'{v}en radius and material is accreted onto the WD via magnetic field lines. In polars the Alf\'{v}en radius is large enough that no disc is formed and matter is fed directly onto the WD's magnetic field lines.

The non-magnetic CVs are further divided into subclasses based on their long-term photometric behaviour. The accretion discs in some CVs undergo a thermal-viscous instability, which switches the disc between a low, faint state and a bright, hot state (\citealt{Smak1971, Osaki1974, Hoshi1979}). The bright states are known as dwarf nova outbursts and are 2--8 mag brighter at optical wavelengths than in the quiescent state. Similar outbursts are seen in X-ray binaries and the same mechanism is believed to be responsible \citep{Lasota2001}. The dwarf nova outbursts typically last for a few days and recur on timescales of weeks to years. CVs that show such outbursts are known as dwarf novae (DN). Systems with a mass-transfer rate from the secondary that is high enough to maintain the accretion disc in a constant hot state are known as (non-magnetic) novalikes. Note that polars and intermediate polars are also sometimes referred to jointly as magnetic novalike systems.

CVs are well known for their variable optical, ultraviolet and X-ray emission, but their radio emission is less well studied. A few studies were performed in the 1980s, but their detection rates were low and the detections were unpredictable. Only three non-magnetic CVs (SU UMa, EM Cyg and TY Psc) were detected at radio wavelengths \citep{Benz1983, Benz1989, Turner1985}. Subsequent re-observations of the same sources with better sensitivities were usually not successful (typically 0.1\,mJy upper-limits) and simply added to the large number of non-detections \citep{Benz1983, Cordova1983, Fuerst1986, Echevarria1987, Nelson1988}.

Similarly, very few of the magnetic CVs have been detected at radio wavelengths, and of those detected, only AM Her, AR UMa and AE Aqr are persistent radio emitters (see \citealt{Mason2007}). The radio emission from the detected polars and intermediate polars (AM Her, V834 Cen, ST LMi, AR UMa, AE Aqr, DQ Her and BG CMi) was highly variable and often showed flares (e.g. \citealt{Abada-Simon1993, Wright1988, Pavelin1994, Mason2007}). AM Her in particular, showed strong flares and even variable circular polarization that peaked at 100\% during a 9.7 mJy flare \citep{Dulk1983, Chanmugam1987}. The emission mechanism is not known, but in individual sources it has been suggested to be non-thermal emission from gyrosynchrotron or cyclotron maser radiation (e.g. \citealt{Mason2007, Meintjes2005}).

The lack of radio detections had implications for both CV studies and accretion theory. As radio emission is often taken as a tracer for jets, it was accepted that CVs do not launch jets, and this was used to constrain jet launching models \citep{Livio1999, Soker2004}. CVs would thus be the only accreting compact objects to not launch jets, as jets have been found in other compact accreting binaries, including those containing WDs (super soft sources, symbiotics and novae).

After a long hiatus in radio observations of non-magnetic CVs, \citet{Koerding2008} and \citet{Miller-Jones2011} obtained radio light-curves during outbursts of the prototypical dwarf nova SS Cyg. It showed a bright radio flare ($>$ 1 mJy) at the start of the outburst, followed by weaker radio emission (0.3-0.1 mJy) during the plateau phase of the optical outburst. This pattern was observed in multiple outbursts and a direct measurement of the distance to SS Cyg was determined using the radio parallax method \citep{Miller-Jones2013}. In light of these detections, one can understand the earlier non-detections. Given the comparatively low sensitivity of radio telescopes at the time, the earlier observations needed to catch the flare by chance, as the plateau emission would have been undetectable. This may have been the case with EM Cyg \citep{Benz1989}. 

Besides this first secure detection of an outbursting (DN-type) CV, a nonmagnetic novalike CV has also been detected. \citet{Koerding2011} detected the novalike V3885 Sgr at 6 GHz (C-Band) at a flux density of 0.16 mJy (distance of 110$\pm$30 pc; \citealt{Hartley2002}). This flux is consistent with that of SS Cyg during the outburst plateau phase (0.15 to 0.5 mJy at 8.5 GHz), which given the similar distance (114 pc; \citealt{Miller-Jones2013}) implies a similar radio luminosity.

To establish the emission mechanism (or mechanisms) of the radio emission observed in non-magnetic CVs, one needs a larger sample of radio-detected CVs - particularly at higher sensitivity. In this paper we present the results of a pilot survey of 4 additional novalike cataclysmic variables conducted with the upgraded Very Large Array (VLA).

In Section \ref{sec:targets} we present our targets. The VLA observations and data reduction are described in Section \ref{sec:observations}. In Section \ref{sec:results} we present the results and discuss them in Section \ref{sec:discussion}.

%%%%%%%%%%%%%%%%%%%%%%%%%%%%%%%%%%%%%%
\section{Targets}\label{sec:targets}

We selected the four nearest and brightest novalike CVs from the \citet{Ritter2003} catalogue that are observable with the Karl G. Jansky Very Large Array (VLA). Preferentially we targeted non-magnetic novalikes, but also included SW Sex type novalikes, whose peculiar properties have been suggested to be associated with dynamically significant magnetic fields associated with their WDs.

The targets are RW Sex, V1084 Her, TT Ari and V603 Aql. Their V-band magnitudes and best distance estimates are given in Table \ref{tbl:coords}. Each source is described briefly below so that the source properties can be compared with the radio observations in Section \ref{sec:results}.

\begin{table*}
  \centering
  \begin{minipage}{140mm}
    \caption{Properties of the target novalikes}
    \begin{tabular}{lllll}
    \hline
    Name & RA (J2000) & Dec. (J2000) & V-mag$^{d}$ & Distance (pc)$^{e}$\\ 
    \hline
    RW Sex & 10:19:56.62309$\pm0.00201^a$ & -08:41:56.0867$\pm0.00156$ & $\sim$10.7 & 150$\pm$37$^{i,h,m}$\\
    V1084 Her & 16:43:45.70$\pm0.07^b$ & +34:02:39.7$\pm0.06$ & $\sim$12.4 & 305$\pm$137$^{j,h,m}$\\
    TT Ari & 02:06:53.084$\pm0.02^c$ & +15:17:41.81$\pm0.026$ & $\sim$10.7 & 335$\pm$50$^{g,k}$\\
    V603 Aql & 18:48:54.63615$\pm0.00223^a$ & +00:35:02.865$\pm0.00182$ & $\sim$12 & 249+9-8$^{f,l}$\\
    \hline
    \multicolumn{5}{p{14cm}}{\footnotesize{\textit{Notes:} Optical coordinates retrieved via Simbad, from $^{a}$\citet{vanleeuwen2007}, $^{b}$\citet{Cutri2003} and $^{c}$\citet{Hog2000}. $^{d}$V-mag at the time of the VLA observations, estimated from long-term AAVSO lightcurves. $^{e}$For each system, we adopted the best available distance estimate. These estimates were based on $^f$the observed parallax, $^g$the direct detection of the primary/secondary, $^h$the absolute infrared magnitude or the H$\alpha$ equivalent width, in that order. Distances from $^i$\citet{Beuermann1992}, $^j$\citet{Ak2008}, $^k$\citet{Gaensicke1999} and $^l$\citet{Harrison2013}. $^m$As distance errors were not quoted for RW Sex and V1084 Her, we have made conservative error estimates based on the distance determination methods used (45\% and 25\% respectively).}}\\
    \end{tabular}
    \label{tbl:coords}
  \end{minipage}
\end{table*}

\subsection*{RW Sex}

RW Sextantis (RW Sex) is an extremely bright novalike, as it has an apparent magitude of $m_V\sim$10.6 mag\footnote{according to the long-term AAVSO light curve} in the $V$-band and absolute magnitude of $M_V$=4.8 mag \citep{Beuermann1992}. \citeauthor{Beuermann1992} determined it to have an orbital period of 0.24507$\pm$0.00020 d, an inclination of 28$\degr$ to 40$\degr$ and a mass ratio of $q=\frac{M_2}{M_1}$=0.74$\pm$0.05 (where $M_1$ and $M_2$ are respectively the mass of the WD and secondary star). RW Sex is known to have a fast disc wind (up to 4500 km s$^{-1}$; e.g. \citealt{Prinja1995,Prinja2003}). Estimates for the mass-transfer rate from the secondary ($\dot{M_2}$) range between $10^{-9}\,M_{\odot}\,\textrm{yr}^{-1}$ and $10^{-8}\,M_{\odot}\,\textrm{yr}^{-1}$ \citep{Linnell2010, Vitello1993, Greenstein1982}.

\citet{Cordova1983} observed RW Sex in the radio with the VLA at 4885 MHz and a bandwidth of 50 MHz. This yielded a non-detection with an upper-limit of 0.15 mJy. No further radio observations were taken until now.

\subsection*{V1084 Her}
V1084 Herculis (V1084 Her; RX J1643.7+3402) is a bright ($m_V\sim12.6$), low-inclination novalike with an orbital period of 0.12056$\pm$0.00001 d \citep{Mickaelian2002, Patterson2002}. It is at a distance of $\sim$305 pc \citep{Ak2008} and is classified as a SW Sex type novalike \citep{Mickaelian2002, Patterson2002}.

There are a number of properties that define the SW Sex class, but the dominant characteristic is that the emission lines are single-peaked, despite the (mostly) high-inclination accretion discs -- see \citet{Rodriguez-Gil2007} for an overview. Individual members of this class have shown evidence for having magnetic white dwarf primaries (e.g. \citealt{Rodriguez-Gil2009,Rodriguez-Gil2001,Rodriguez-Gil2002,Baskill2005}), but it is not clear that this is the case for every SW Sex star (e.g. \citealt{Dhillon2013}). V1084 Her is not classified as an intermediate polar, but it has been argued that the WD in V1084 Her is magnetic \citep{Rodriguez-Gil2009}. This is based on their discovery of modulated optical circular polarization at a period of 19.38$\pm$0.39 min and emission line flaring at twice this polarimetric period.

V1084 Her has not been observed at radio wavelengths prior to this study.

\subsection*{TT Arietis}
TT Arietis (TT Ari) has been observed and studied routinely since its discovery \citep{Strohmeier1957}. It has an inclination of roughly $30\degr$ \citep{Cowley1975}, an orbital period of 0.13755040$\pm$0.00000017 d \citep{Wu2002}, a 39,000 K WD and a M3.5 type secondary star \citep{Gaensicke1999}. A rough estimate for the mass-accretion rate ($\dot{M_1}$) is 2.8--26.7$\times10^{-8}M_{\odot}\,\textrm{yr}^{-1}$ \citep{Retter2000} and far-ultraviolet observations indicate the presence of a fast and variable disc wind \citep{Prinja1995,Hutchings2007}.

Due to its long-term behaviour in the optical\footnote{See the AAVSO light curve at http://www.aavso.org/data/lcg}, TT Ari is classified as a VY Sculptoris (VY Scl) type novalike (see \citealt{Shafter1985}), as it spends most of the time in a high-state ($m_V\sim$12--14 mag), but shows occasional low states ($m_V>19$ mag) lasting a few hundred days.

Previously it was argued that TT Ari is an intermediate polar. First, it shows a high X-ray luminosity and variability \citep{Robinson1994}. Second, the photometric period -- which differed from the spectroscopic period (\citealt{Tremko1992} and \citealt{Robinson1994}) -- was incorrectly taken as the spin period of the WD. It was subsequently shown that the photometric period was produced by a negative superhump (e.g. \citealt{Vogt2013}) and that the X-ray properties of TT Ari fit well into the properties of non-magnetic CVs \citep{vanTeeseling1996}, thereby establishing TT Ari as a non-magnetic system.

Although TT Ari has been well studied at optical, X-ray and UV wavelengths, only one observation was taken in the radio. \citet{Cordova1983} observed it during an optical low-state with the VLA at 4885 MHz (50 MHz bandwidth), but did not detect it. The upper-limit they obtained was 0.44 mJy. No further radio observations were taken of it until now.

\subsection*{V603 Aql}

V603 Aql is Nova Aquilae 1918 -- the brightest nova eruption (thermo-nuclear runaway on the surface of the white dwarf) of the 20th century. The eruption began on June 4, 1918, peaked 6 days later at $m_V=-$0.5 mag and was back at pre-nova brightness ($m_B$=11.43$\pm$0.03 mag) by February 1937 \citep{Strope2010,Johnson2014}. Since the eruption, it has been fading by 0.44$\pm0.04$ mag century$^{-1}$ \citep{Johnson2014}; novae have been predicted to fade after outburst, as the outburst should widen the binary and consequently pause mass transfer (see e.g. \citealt{Shara1986} and \citealt{Patterson2013}). By June 1982 only a very faint shell was still visible \citep{Haefner1985}.

The system parameters include an orbital period of 0.1385$\pm$0.0002 d, inclination of 13$\degr\pm$2$\degr$, white dwarf mass $M_1$=1.2$\pm0.2M_{\odot}$ and mass ratio $q$=0.24$\pm$0.05 \citep{Arenas2000}. It shows a time-variable wind and the mass accretion rate is estimated to be around $\dot{M_1}$=9.2--94.7$\times10^{-9}M_{\odot}\,\textrm{yr}^{-1}$ \citep{Retter2000, Prinja2000}. 

As in the case of TT Ari, it was argued that V603 Aql could be an intermediate polar. This stemmed from detections of X-ray periodicities, linear and circular polarization and a differing photometric and spectroscopic period (e.g. \citealt{Haefner1985, Gnedin1990, Udalski1989}). Since then it has been confirmed that the photometric period is the permanent superhump period and that it shows no coherent sub-orbital period oscillations -- thereby establishing V603 Aql as a non-magnetic CV \citep{Patterson1991, Patterson1993, Naylor1996, Patterson1997, Borczyk2003, Andronov2005, Mukai2005}. Furthermore, \citet{Mukai2005} state that V603 Aql does not show a strong energy dependence in X-ray variability, unlike what is seen in IPs.

No radio observations have been taken of V603 Aql prior to this study.

%%%%%%%%%%%%%%%%%%%%%%%%%%%%%%%%%%%%%%
\section{Observations}\label{sec:observations}

\begin{table*}
  \centering
  \begin{minipage}{14cm}
    \caption{Observing log}
    \begin{tabular}{|l|p{0.6cm}|p{2cm}|p{2cm}|p{3cm}|p{2cm}|p{2cm}|}
    \hline
    Name & Obs. No. & Date and Start Time (UT) & Integration Time (s) & Bandpass, Flux and Polarization Angle calibrator & Amplitude and Phase Calibrator & Polarization Leakage Calibrator\\ %[0.5ex]
    \hline
    RW Sex & 1 & 13/03/2014 08:17:09.0 & 2268 & 3C286 & J0943-0819 & J1407+2827 \\ %[1ex]
    & 2 & 15/03/2014 07:57:21.0 & 2264 & 3C286 & J0943-0819 & J1407+2827 \\
    V1084 Her & 1 & 22/03/2014 07:40:37.0 & 2364 & 3C286 & J1635+3808 & J1407+2827 \\
    & 2 & 31/03/2014 14:20:34 & 2126 & 3C286 & J1635+3808 & J1407+2827\\
    TT Ari & 1 & 02/04/2014 00:08:24 & 2138 & 3C48 & J0203+1134 & J0319+4130\\
    & 2 & 02/04/2014 18:56:25.0 & 2304 & 3C48 & J0203+1134 & J0319+4130\\
    V603 Aql & 1 & 07/04/2014 13:52:33.0 & 2144 & 3C48 & J1851+0035 & J2355+4950\\
    & 2 & 14/04/2014 14:09:57.0 & 2144 & 3C48 & J1851+0035 & J2355+4950\\
    \hline    
    \multicolumn{7}{p{14cm}}{\footnotesize{Observations were taken at 4226--8096 MHz (in C-band) with 4096 MHz of bandwidth (3-bit mode), in the VLA A-configuration.}}\\
    \end{tabular}
    \label{tbl:log}
  \end{minipage}
\end{table*}

Two separate 1-hour observations with the VLA were taken of each target, both to look for variability and to facilitate easier scheduling. The log of the observations is given in Table \ref{tbl:log}.

The observations were taken in the most extended A-configuration (baselines ranging from 0.68 to 36 km) at $4-8\,$GHz (C-band), with 4 GHz bandwidth (3-bit samplers) for the highest sensitivity. This decision was based on the steep spectrum (spectral index of --0.7) of the only novalike for which the spectral index in the radio was known, namely V3885 Sgr \citep{Koerding2011}. The frequency tuning of the receivers was shifted to 4226--8096 MHz in order to avoid RFI from the Clarke belt. The 4 GHz bandwidth was split into two basebands, 4226--6246 MHz and 6176--8096 MHz, each of which was divided into 16 spectral windows comprising 64 channels. The observations were taken in standard phase referencing mode, where the phase calibrator was observed for 1 minute every 8 minutes.

The data were reduced with the VLA calibration pipeline v1.2.0 and the polarization calibration was performed in CASA\footnote{Common Astronomy Software Applications package \citep{McMullin2007}} v4.2.0 following standard procedures. The absolute flux density scale was set via observations of a standard flux calibrator source (3C286 or 3C48; see Table 2), using the Perley-Butler 2010 coefficients within CASA. No self-calibration was performed. For the imaging, two Taylor terms were used to model the frequency dependence, and Briggs weighting with robust parameter 1 was chosen to suppress bright source sidelobes and improve the resolution. Fitting was performed in the image plane using imfit task within CASA, and the RMS noise level for each image was calculated in the vicinity of the target.  

%%%%%%%%%%%%%%%%%%%%%%%%%%%%%%%%%%%%%%
\section{Results}\label{sec:results}

\begin{table*}
  \centering
  \begin{minipage}{14cm}
    \caption{Results}
    \begin{tabular}{|l|l|r|l|l|r|l|c|l|}
    \hline
    Name, Obs. & Beam Size$^a$ & PA$^b$ & RA Offset$^c$ & DEC Offset$^c$ & Peak Flux$^d$ & RMS$^d$ & CP$^d$ & LP$^d$\\
    \hline
    RW Sex, 1 & $0.58''\times0.33''$ & $39\degr$ & 0.016$\pm$0.02 & 0.26$\pm$0.02 & 33.6 & 3.7 & $<$13.8 & $<$7.2\\
    RW Sex, 2 & $0.56''\times0.33''$ & $36\degr$ & 0.018$\pm$0.02 & 0.18$\pm$0.03 & 26.8 & 3.3 & $<$12.9 & $<$8.4\\
    V1084 Her, 1 & $0.43''\times0.31''$ & $-71\degr$ & - & - & $<$10.2 & 3.4 & - & -\\
    V1084 Her, 2 & $0.44''\times0.32''$ & $80\degr$ & - & - & $<$11.4 & 3.8 & - & -\\    
    TT Ari, 1 & $0.67''\times0.30''$ & $64\degr$ & 0.015$\pm$0.03 & 0.33$\pm$0.02 & 39.6 & 4.2 & 27.4$\pm$4.2 & $<$8.1\\
    TT Ari, 2 & $0.33''\times0.30''$ & $0\degr$ & 0.013$\pm$0.003 & 0.300$\pm$0.004 & 239.1 & 5.5 & 22.8$\pm$4.8 & $<$9.6\\
    V603 Aql, 1 & $0.40''\times0.38''$ & $31\degr$ & 0.010$\pm$0.003 & 0.095$\pm$0.003 & 178.2 & 4.3 & $<$12.9 & $<$8.7\\
    V603 Aql, 2 & $0.46''\times0.33''$ & $33\degr$ & 0.010$\pm$0.003 & 0.085$\pm$0.004 & 190.5 & 3.9 & $<$12.0 & $<$8.1\\
    \hline    
    \multicolumn{9}{p{14cm}}{\footnotesize{$^a$Major and minor axis of the synthesized beam. $^b$Position angle of the beam. $^c$Offset (in arcsec) from the optical position given in Table \ref{tbl:coords}. $^d$Units are $\mu$Jy beam$^{-1}$. All detections were consistent with point sources. Upper-limits are quoted as $3\sigma$.}}\\
    \end{tabular}
    \label{tbl:results}
  \end{minipage}
\end{table*}

\subsection{Total fluxes}

Three out of the four systems were detected with $\geq 8\sigma$ significance -- a 75\% detection rate. RW Sex, TT Ari and V603 Aql were detected in both their observations, while V1084 Her was detected in neither. None of the sources were resolved (see Fig. \ref{fig:stokesI}) and their measured radio positions are consistent with the optical positions. These results are summarized in Table \ref{tbl:results}.

\subsection{Polarization}

We also imaged the target fields in all four Stokes parameters (I, Q, U and V) to determine the source polarization. TT Ari was the only source where circular polarization was measured and none of the sources were linearly polarized. Both observations of TT Ari showed circular polarization (CP, see Table \ref{tbl:results} and Figure \ref{fig:ttari_stokesV}). The fractional polarization in observation 1 was 70\%, whereas in observation 2 it was $\sim10$\%. This difference in fractional polarizations was due to a large increase in unpolarized flux between the observations, as the polarized fluxes for the two observations were similar (at 27.4$\pm$4.2 and 22.8$\pm$4.8 $\mu$Jy beam$^{-1}$).

As the VLA has circular feeds, it is possible to produce instrumental CP, extrinsic to the source. We therefore ran a number of checks to test if this was the case. None of the other sources in the field show CP (the closest source is less than 2 arcsec from TT Ari and the brightest source is 58 $\mu$Jy beam$^{-1}$). Neither the flux nor the gain calibrator show CP above 0.3\% of the total intensity. Imaging the field with half of the antennas yields the same level of polarization as for the other half, so it it unlikely to be an antenna calibration problem. Similarly, when the two basebands are imaged separately they both show CP, so it is not a single-baseband effect. In the observational setup, the targets were offset 5 arcsec from the phase center to avoid correlator artifacts at the phase centre. Finally, both observations of TT Ari show circular polarization and none of the other novalikes (which were all observed with the same setup) showed CP. We therefore conclude that the CP is likely to be intrinsic to TT Ari and not an instrumental or calibration artifact.

\subsection{Variability}

\begin{figure*}
  \centering
    \includegraphics[width=140mm]{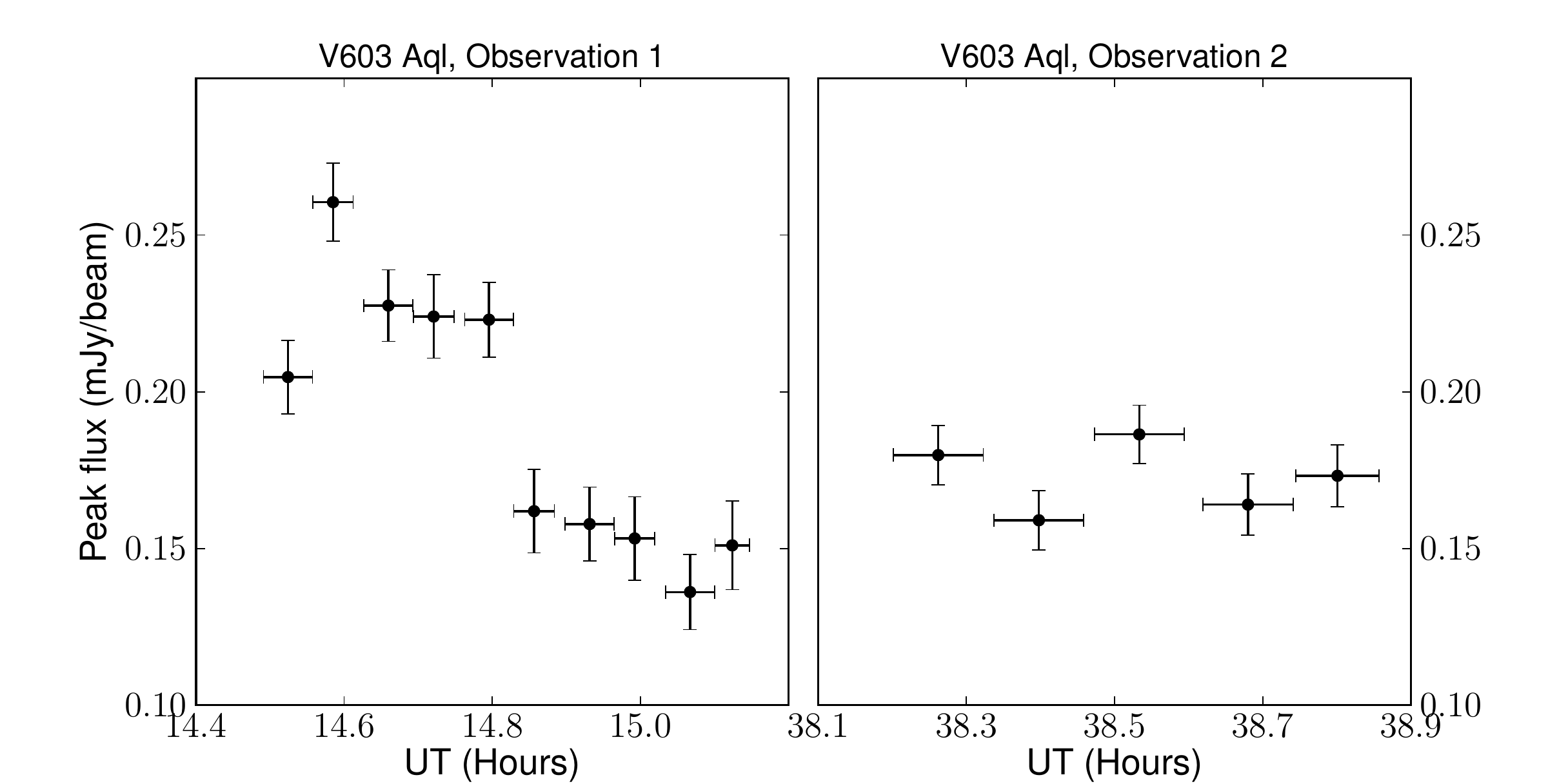}
    \caption{Total intensity (Stokes I) light curve for observations 1 and 2 of V603 Aql. There is clear variability in observation 1, but no such variability in observation 2. Error bars on the x-axis show the integration time for each point.}
  \label{fig:v603aql_variability}
\end{figure*}

TT Ari and V603 Aql were both variable on a timescale of minutes.

The total intensity light curve for V603 Aql is given in Figure \ref{fig:v603aql_variability}. It peaked at 260.5$\pm12.5\,\mu$Jy beam$^{-1}$ during the first half of observation 1, but then dropped to approximately 170 $\mu$Jy beam$^{-1}$ for the second half. We could detect variability on time-scales down to 217 s. In the second observation, V603 Aql was not variable.

\begin{figure*}
  \centering
    \includegraphics[width=\textwidth]{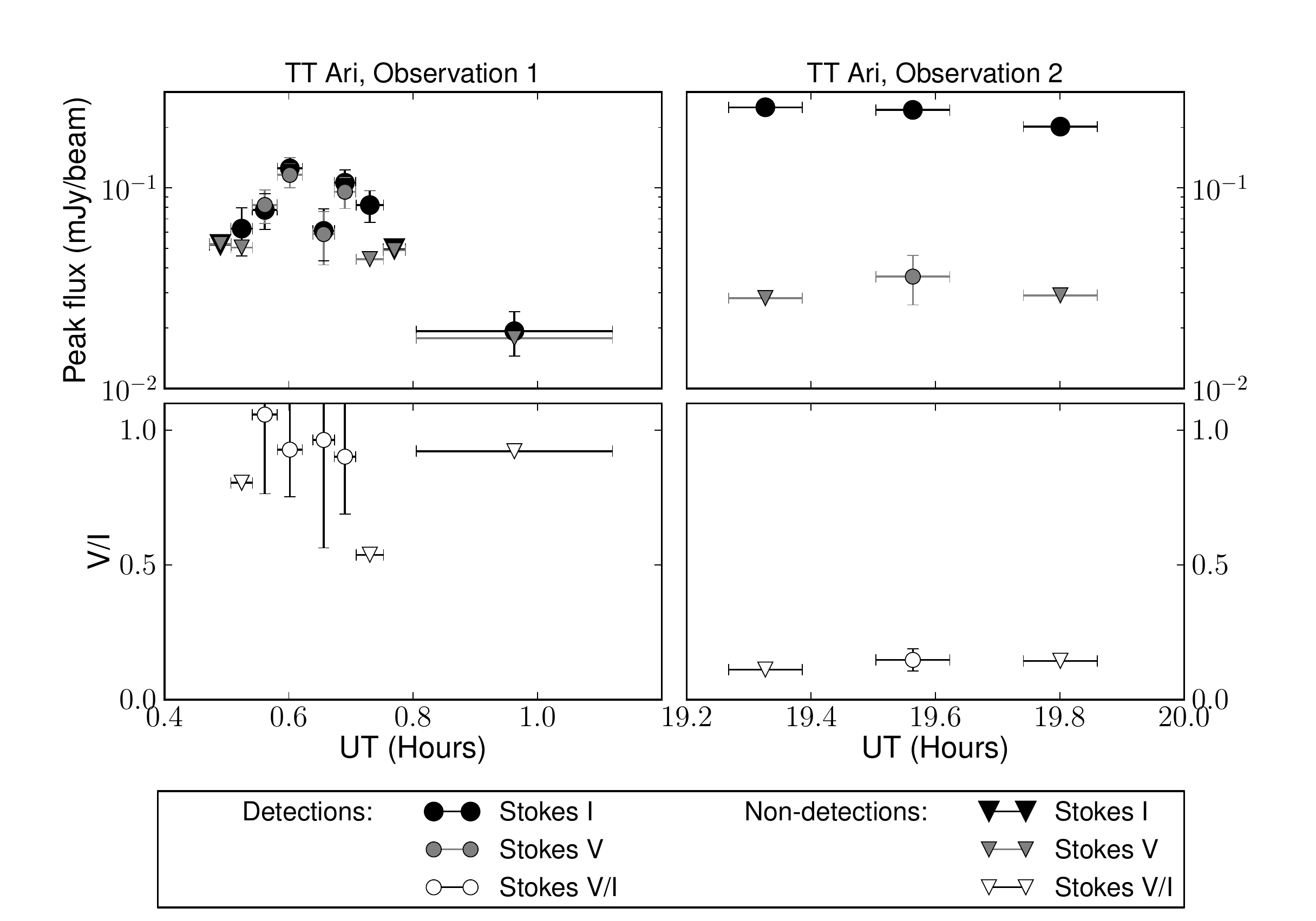}
    \caption{Total intensity (Stokes I) and circular polarization (Stokes V) light curves for observations 1 and 2 of TT Ari. The strong flare in Observation 1 of TT Ari is consistent with 100\% circular polarization. 1-Sigma error bars are shown for the peak flux; they are too small to be seen in observation 2. The error bars on the x-axis give the integration time for each point.}
  \label{fig:ttari_variability}
\end{figure*}

In the 19 hours between observations 1 and 2 of TT Ari, its flux increased by a factor of 6 (39.6$\pm$4.2 to 239.1$\pm$5.5 $\mu$Jy beam$^{-1}$). Splitting the observations up in time (see Figure \ref{fig:ttari_variability}) shows that the variability is actually on shorter timescales (detected down to 144 s), and most of the flux from observation 1 was detected during a $\sim10$-min period. All the circularly polarized emission arose from a $\sim$8 min flare during this time and the fractional polarization reached up to 100\%. The flux dropped to 19.3$\pm4.8\,\mu$Jy beam$^{-1}$ in the second half of the observation.

Observation 2 was also variable (Figure \ref{fig:ttari_variability}), but at a higher total flux (201.8$\pm9.8$ -- 251.9$\pm9.4\,\mu$Jy beam$^{-1}$) and a lower CP fraction ($\leq$15\%). As mentioned previously, the polarized flux for the two observations was similar (see Table \ref{tbl:results} and Figure \ref{fig:ttari_stokesV}).

Since the other sources in the field (in both cases) were not variable, we conclude this variability is intrinsic to TT Ari and V603 Aql.

The flux density of the first and second obervation of RW Sex were consistent, but we checked for shorter time-scale variability by splitting the two observations into four epochs (each $\sim$20 minutes long) and imaging each separately. As the flux densities were consistent in all four epochs, we conclude the RW Sex is not variable.

V1084 Her was not detected in either observation, but if it showed only a minor, short-duration flare (similar to TT Ari), this would not be detectable when integrated over the whole observation. In order to test if this was the case, we imaged the first and second halves of the two observations separately. We found no detections down to $3\sigma$ upper-limits of $\sim17.5\,\mu$Jy beam$^{-1}$ on $\sim$20 minute timescales.

\subsection{Spectral indices}

\begin{table*}
  \centering
  \begin{minipage}{140mm}
    \caption{Spectral indices}
    \begin{tabular}{|l|l|l|l|l|l|l|l|}
    \hline
    Object & Observation & Stokes & Spectral Index & Reduced $\chi^2$ & Band & Peak Flux & RMS\\
    &  &  & (F=$\nu^{\alpha}$) &  & (MHz) & ($\mu$Jy beam$^{-1}$) & ($\mu$Jy beam$^{-1}$)\\
    \hline
    RW Sex & 1 and 2$^a$ & I & -0.5$\pm$0.7 & - & 4226--6274 & 36.2 & 5.4\\
    & & & & & 6176--8224 & 30.5 & 4.6\\
    \hline
    TT Ari & 1 & I & 1.7$\pm$0.8 & - & 4226--6274 & 28.6 & 6.2\\
    & & & & & 6176--8224 & 49 & 5.8\\
    \hline
    TT Ari & 1, during flare$^b$ & I & 1.6$\pm$0.1 & 0.05 & 4226--5250 & 59.7 & 16.4\\ 
    & & & & & 5250--6274 & 74.0 & 19.2\\
    & & & & & 6176--7200 & 95.0 & 16.3\\
    & & & & & 7200--8224 & 123.0 & 11.3\\
    \hline
    TT Ari & 1, during flare$^b$ & V & 1.31$\pm$0.06 & 0.01 & 4226--5250 & 56.7 & 16.0\\
    & & & & & 5250--6274 & 73.3 & 18.7\\
    & & & & & 6176--7200 & 85.9 & 16.3\\
    & & & & & 7200--8224 & 106.9 & 11.1\\
    \hline
    TT Ari & 2 & I & 0.7$\pm$0.3 & 4.9 & 4226--5250 & 173.4 & 10.5\\
    & & & & & 5250--6274 & 240.2 & 12.0\\
    & & & & & 6176--7200 & 264.9 & 11.8\\
    & & & & & 7200--8224 & 258.0 & 11.2\\
    \hline
    V603 Aql & 1 & I & 0.54$\pm$0.05 & 0.15 & 4226--5250 & 152.7 & 7.7\\
    & & & & & 5250--6274 & 173.5 & 8.7\\
    & & & & & 6176--7200 & 189.6 & 9.5\\
    & & & & & 7200--8224 & 199.1 & 7.2\\
    \hline
    V603 Aql & 2 & I & 0.16$\pm$0.08 & 0.2 & 4226--6274 & 178.9 & 8.9\\
    & & & & & 5250--6274 & 192.4 & 10.7\\
    & & & & & 6176--7200 & 189.0 & 15.0\\
    & & & & & 7200--8224 & 193.0 & 14.7\\
    \hline    
    \multicolumn{8}{p{14cm}}{\footnotesize{Notes: The spectral indices were obtained by fitting a power law to the measurements in different frequency sub-bands. $^a$Observation 1 and 2 were combined to reduce the uncertainty in the spectral index. $^b$In the time range 00:32:30--00:42:28, during which CP was detected in the flare. The errors used in the fit were the RMS values.}}\\
    \end{tabular}
    \label{tbl:spectral_index}
  \end{minipage}
\end{table*}

\begin{figure}
  \centering
    \includegraphics[width=80mm]{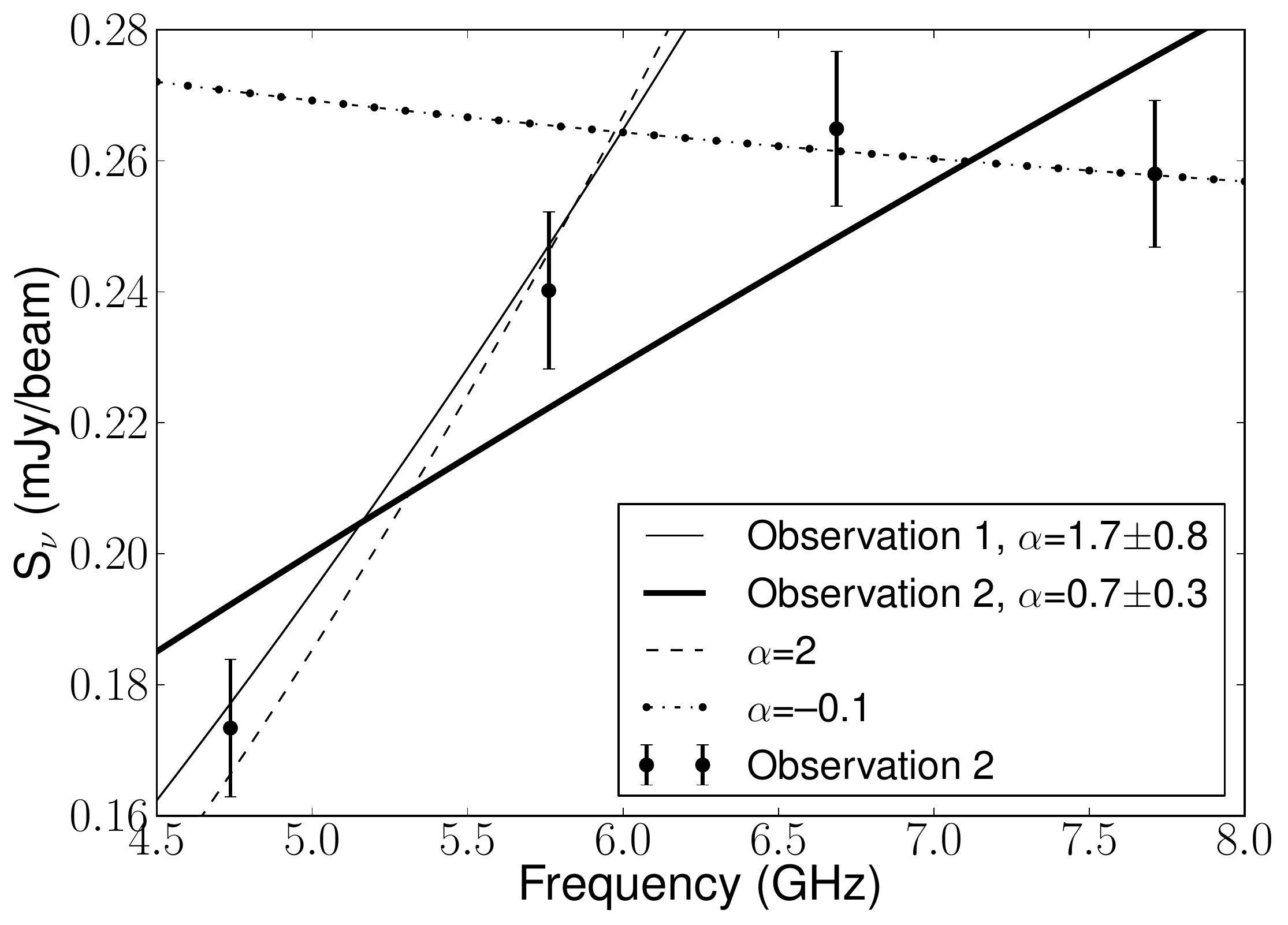}
    \caption{Spectrum of the second observation of TT Ari. The spectrum cannot be fitted with a single spectral index. The spectral index from the first observation, $\alpha$=2 and $\alpha$=--0.1 are shown for comparison.}
  \label{fig:ttari_spectral}
\end{figure}

The observations showed a spread of spectral indices (Table \ref{tbl:spectral_index}), from $-0.5\pm$0.7 to 1.7$\pm$0.8.

As the total flux in the first observation of TT Ari was dominated by the $\sim$10 minute flare, it is not suprising that the spectral index taken during the flare ($\alpha=1.6\pm0.1$) is consistent with that taken over the whole of the observation. Unfortunately there was insufficient signal-to-noise to determine the spectral index after the flare. The circular polarization (which was only detected during the flare) had $\alpha=1.31\pm0.06$. The second, brighter observation was not fitted well with a single power law, but rather showed a spectral turnover -- this is plotted in Figure \ref{fig:ttari_spectral}.

The spectrum of V603 Aql flattened from $\alpha=0.54\pm0.05$ in the first observation to 0.16$\pm$0.08 in the second (fainter) observation.

%%%%%%%%%%%%%%%%%%%%%%%%%%%%%%%%%%%%%%
\section{Discussion}\label{sec:discussion}

Historically, non-magnetic CVs have not been considered to be significant radio emitters. This stems from the low detection rates in previous surveys. In the 1980s, more than 50 radio observations of non-magnetic CVs were taken and, as summed up by \citet{Benz1996}, only two were detected, and only twice each. In contrast, we have obtained a 75\% detection rate in this survey -- strongly indicating that many novalikes are indeed significant radio emitters and that with modern radio telescopes we have the sensitivity required to detect them.

\begin{figure*}
  \centering
    \includegraphics[width=\textwidth]{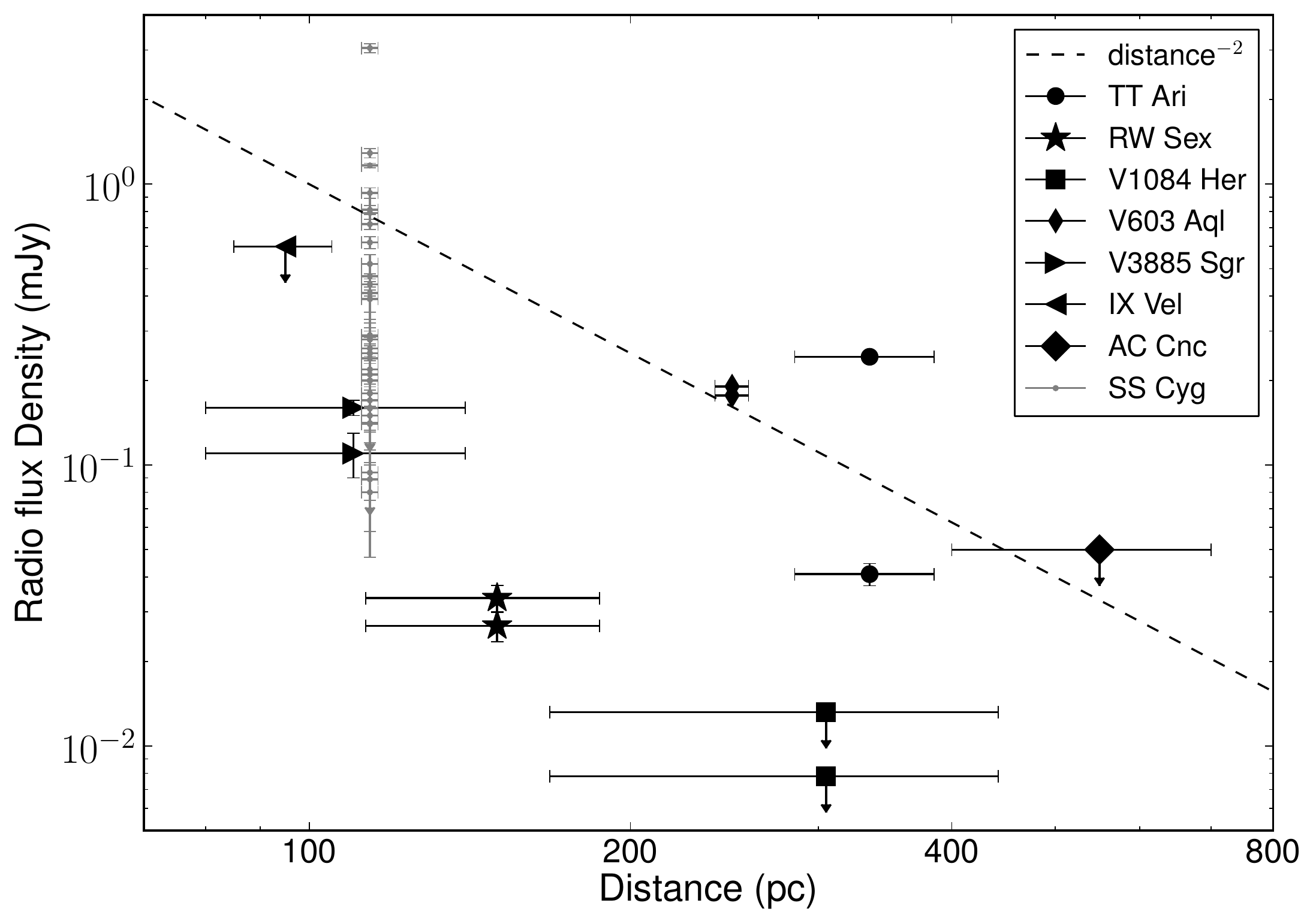}
    \caption{Radio flux density of all high-sensitivity observations of non-magnetic CVs, taken since 2008, as a function of distance (this work, \citealt{Koerding2008,Koerding2011,Miller-Jones2011,Miller-Jones2013}). The dotted line shows the expected trend ($1/d^2$) for sources with equal luminosities. Errors are calculated via standard error propagation techniques. Observations taken of the dwarf nova SS Cyg at various stages of outburst are plotted for comparison.}
    \label{fig:radiodistance}
\end{figure*}

All the observations of non-magnetic CVs conducted since 2008 are plotted in Fig. \ref{fig:radiodistance}. The radio fluxes are below the $\sim$0.1 mJy detection limits of previous radio surveys.

\begin{figure*}
  \centering
    \includegraphics[width=\textwidth]{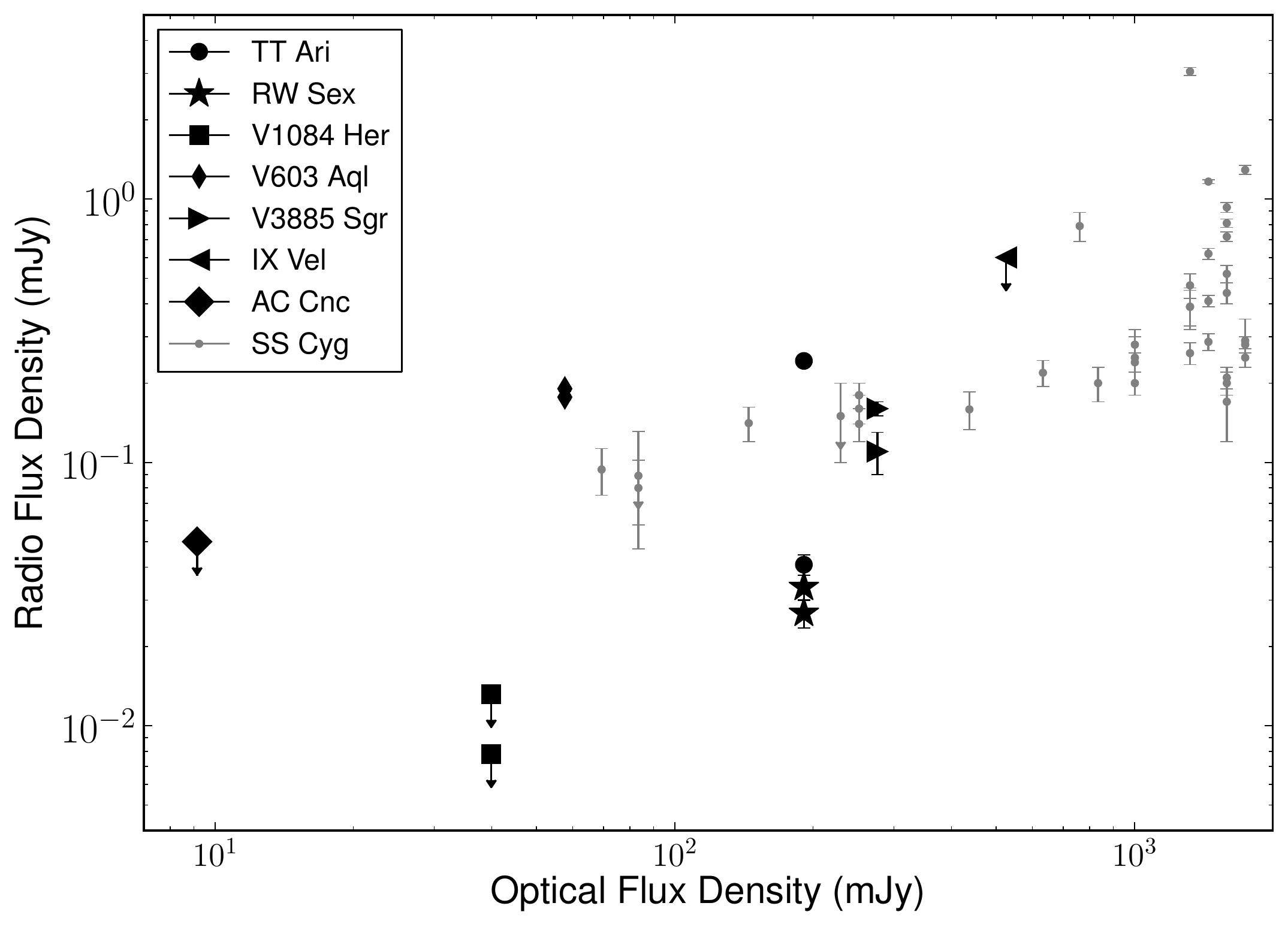}
    \caption{Radio and optical fluxes of all high-sensitivity (recent) detections and non-detections of non-magnetic CVs -- with the dwarf nova SS Cyg plotted for comparison \citep{Koerding2008,Koerding2011,Miller-Jones2011,Miller-Jones2013}. Using a rank test (Kendell's tau), we find that there is no significant correlation between radio and optical fluxes for the given data (p value of 0.7)}
  \label{fig:radiooptical}
\end{figure*}

There are not enough detections to test if there is a correlation between the radio flux and distance, as expected for a sample of uniform luminosity. Consequently, and due to the large distance uncertainty, we cannot say whether V1084 Her is simply too far away to detect or if it is intrinsically faint. At this stage there is also no correlation between the radio and optical fluxes of the novalikes (see Fig. \ref{fig:radiooptical}). The dwarf nova SS Cyg does show a general positive correlation in radio flux with optical flux, but this is emission from a radio flare at the start of outburst, after which the radio flux was undetectable (Fig. 2 in \citealt{Koerding2008, Miller-Jones2011}).

The range of different fluxes, variability, circular polarizations and spectra observed in this sample indicate there is likely more than one emission mechanism at work. Several explanations for the origin of the radio emission in non-magnetic CVs have been proposed over the years. These include thermal emission, synchrotron (possibly from jets) or gyrosynchtron emission, and cyclotron maser emission.

\subsection{Thermal Emission}\label{sec:disc_thermal}

Thermal emission could be produced by a large gas cloud surrounding the DN that is formed by the wind during outburst (e.g. \citealt{Cordova1983,Fuerst1986}). For the dwarf nova SS Cyg this suggestion was ruled out due to the observed brightness temperature, spectrum and coincidence with the optical outburst \citep{Koerding2008}. Thermal emission could produce the observed spectral indices in our sample.

As all of our detections are unresolved, we can place an upper-limit on the size of the emitting region. RW Sex is the closest CV and observation 1 had the largest beamwidth (150 pc and 0.58'' respectively); this gives an upper-limit on size of the emitting region of $\sim 1\times10^{15} {\rm cm}$ for our sample.

The brightness temperature of a source is given by
\begin{equation}
 T_b = \frac{S_{\nu}c^2}{2k_B\Omega \nu^2}\\
\end{equation}
where $S_{\nu}$ is the specific flux, $k_B$ is the Boltzmann constant, $\nu$ is the frequency and $\Omega$ is the solid beam angle.

For CVs with orbital periods in the range in this sample, the orbital radius is $r_{\rm{orbit}}\sim10^{11}$ cm. If we assume a circular source (as projected on the sky) that is the size of the binary, this gives a brightness temperature of
\begin{equation}
 T_b \sim 1\times10^{12}\left(\frac{S_{\nu}}{\rm{mJy}}\right)\left(\frac{\nu}{\rm{GHz}}\right)^{-2}\left(\frac{r}{r_{\rm{orbit}}}\right)^{-2}\,\rm{K},
\end{equation}\label{eq:Tb}
where $r$ is the radius of the source.

If the emitting region is the size of the binary, then the brightness temperature for these observations is $\sim1\times10^9$ K. As optically thick thermal emission from an ionized gas typically has brightness temperatures of $10^4-10^5$ K, this implies that any emission of order the size of the orbit must be non-thermal. The emitting region would need to have a radius of $\sim10^2-10^3$ times the orbital radius if the observed emission is optically thick thermal emission, which is unlikely.

In the case of TT Ari and V603 Aql, optically thick thermal emission can be ruled out by the observed variability time-scales and causality arguments. CVs are non-relativistic, so heat transfer occurs at speeds significantly less than the speed of light. CV winds of up to 5000 km s$^{-1}$ have been detected \citep{Kafka2009}, so if we take an exceedingly fast CV wind of $1\times10^4$ km s$^{-1}$, then changes can only be propagated over the binary separation in $\sim$200 s. V603 Aql is variable on time-scales down to 217 s and TT Ari to 144 s, so the emitting region would need to be smaller than the orbit, which (as shown above) cannot be the case for optically thick thermal emission. In addition to these arguments, in the case of TT Ari, thermal emission could also not account for the circular polarization.

The spectrum of observation 2 of TT Ari, however, is suggestive of thermal emission with a turnover at 6 GHz, so we now consider the possibility that there is a thermal component to the radio emission. Figure \ref{fig:ttari_spectral} shows the spectrum; it is consistent with $\alpha$=2 up to 6 GHz and $\alpha$=--0.1 at higher frequencies. This contrasts with the spectrum from observation 1, which was well fit with a single power-law ($\alpha=1.7\pm0.8$, or $\alpha=1.6\pm0.1$ during the flare). The variability and flux density also differed between the two observations, which suggests different emission mechanisms in the two epochs.

We now consider the properties of a possible thermal component in observation 2. For thermal opacity we have that
\begin{equation}
 \tau_{\nu} \sim 8.235\times10^{-2}T_e^{-1.35}\nu^{-2.1}EM, 
\end{equation}
where the frequency $\nu$ is in GHz, $T_e$ is the electron temperature (in K), and $EM$ is the emission measure (pc cm$^{-6}$), which is defined as the integral of the electron number density $n_2$ (in cm$^{-3}$) along the line of sight
\begin{equation}
 EM = n_2^2\rm{d}l\,. 
\end{equation}

For significant ionization $T_e$ must be at least of order 10$^3$ K. If we take $T_e$=5000 K and assume an emitting region that is $Z$ times as large as the orbital radius, then we can estimate the electron density as follows: 
\begin{align}
 EM &\sim \langle n_e^2\rangle Z \left(\frac{r_{\rm{orbit}}}{\rm{pc}}\right) \nonumber\\
 \langle n_e^2 \rangle &\sim 4\times10^7Z^{-0.5}\,\rm{cm}^{-3}
\end{align}

Assuming a spherical emitting region with radius $r\sim1\times10^{14}$ cm (the size restriction based on the brightness temperature) and width $Z$ times the orbital radius (d$r=Zr_{\rm{orbit}}$), we can estimate the total mass of a thermal emitting region as
\begin{align}
 M_t &= 4\pi r^2 n_e m_p\rm{d}r\nonumber\\
 M_t &\sim 8\times10^{23}Z^{0.5}\,\rm{g,}\label{eq:mass_thermal}
\end{align}
where $m_p$ is the mass of a proton (g). If the emission was indeed thermal, $Z$ could be derived by watching the evolution of the radio light curve past epoch 2.

The observed spectrum with $\alpha$=2, and $\alpha$=-0.1 at higher frequencies is more compatible with a thin dense shell (e.g. of a nova) than an extended, centrally concentrated ($r^{-2}$) stellar wind. The latter would have $\alpha$=0.6 at lower frequencies, breaking to $\alpha$=-0.1 and would need a rather contrived geometry in order to reproduce the observed spectrum.

If there is a non-thermal component to the emission in the second observation of TT Ari, then more than one emission mechanism is necessary to produce the observed properties. Consequently we do not favour this scenario. 

\subsection{Non-thermal Emission}\label{sec:disc_nonthermal}

Non-thermal emission from CVs has been suggested by a number of authors (e.g. \citealt{Fuerst1986, Benz1989, Benz1996, Koerding2008}) in the form of gyrosynchrotron and synchrotron emission, and maser emission.

\subsubsection{Gyrosynchrotron Emission}\label{sec:gyrosynchrotron}

\citet{Fuerst1986} concluded that either the magnetic field-strength is insufficient or the production rate of relativistic electrons is too low in non-magnetic CVs to produce gyrosynchrotron radiation, but this conclusion was based on the fact that they did not detect any of the eight non-magnetic CVs they observed at 5 GHz. \citet{Benz1983} had detected EM Cyg prior to this, but \citeauthor{Fuerst1986} were unable to explain this discrepancy. Since then, SU UMa has been detected \citep{Benz1989} and so has V603 Aql (this work), so their conclusion needs revision.

Gyrosynchrotron emission is known to produce highly polarized CP, so it is a plausible emission mechanism for TT Ari. Although the $3\sigma$ upper-limits on the CP fraction in RW Sex and V603 Aql are 12.9\% and 12.0\% respectively, we cannot rule it out for these two sources.

Following the procedure in \citet{Benz1989}, we can estimate the achievable brightness temperature for gyrosynchrotron emission of non-thermal electrons. For typical values of the power-law index of the electrons ($\sim$3) and an average angle between the magnetic field and the line of sight ($\sim60\degr$) we have
\begin{equation}
\mathrm{T}_B < 2.8 \times 10^{8} s^{0.755} \mathrm{K}, 
\end{equation}
where $s$ is the frequency in units of the gyro frequency $eB/m_ec$ and $m_e$ is the electron mass. Thus, if the emission region is limited to the size of the binary, we find that we need a fairly low $s$ factor just above 5, which corresponds to magnetic fields of $B<100\,\mathrm{G}$. As argued by \citet{Benz1989}, one can expect that at least in the case of low magnetic fields together with large emission regions (the size of the orbital separation), the electrons would be adiabatically outflowing and expansion would quench the emission. This could account for the observed flaring behavior in TT Ari.

\subsubsection{Synchrotron Emission}\label{sec:synchrotron}

The spectral index of TT Ari is consistent with optically thick (self-absorbed) synchrotron radiation, but it is circularly polarized and the CP fraction reached 100\% during the flare in observation 1. High levels of linear polarization (LP) are possible, but typically LP levels for synchrotron radiation from astrophysical sources are lower, for example the compact jets in X-ray binaries typically have LP fractions of a few percent (e.g. \citealt{Han1992,Corbel2000,Russell2015}). CP is suppressed in comparison to LP for synchrotron emission of relativistic particles (e.g. \citealt{Longair2011}), so the 100\% CP flare in TT Ari cannot be produced by synchrotron radiation.

V603 Aql and RW Sex both have spectral indices that are consistent with synchrotron radiation. Combined with the lack of strong CP, we could attribute V603 Aql to optically thick synchrotron emission and RW Sex to optically thin synchrotron emission.

\citet{Koerding2008} suggested that radio emission in CVs could be due to synchrotron emission from a jet. This was supported by the observed CV outburst pattern, which has many features in common with X-ray binaries. Jets have been detected in other accreting white dwarfs (symbiotics and novae). In the case of the dwarf nova SS Cyg, \citet{Koerding2008} and \citet{Miller-Jones2011} concluded that the radio emission in outburst was most likely synchrotron radiation produced by a partially quenched optically thick synchrotron jet. This would be consistent with the observed emission from V603 Aql.

\subsubsection{Cyclotron-Maser Emission}\label{sec:maser}

\citet{Benz1989} have suggested that the observed radio emission in CVs is due to a maser, or cyclotron instability, originating in the strong magnetic field and low densities near the white dwarf. They detected variability on time-scales of days and a CP fraction of 81\% from the DN EM Cyg, and concluded that these properties support this model. Maser emission can produce high levels of CP and according to \citet{Benz1996}, best explains the short, sporadic bursts of radio emission detected in DN. Our observations of TT Ari show this type of behaviour.

\citet{Benz1989} estimated the magnetic field strength necessary for EM Cyg to produce cyclotron-maser emission. They assumed a loss-cone velocity distribution for weakly relativistic electrons, the density profile for an isotropic outflow, and a dipole magnetic field for the WD of the form $B(r)=10^5\left(\frac{r_{\rm{WD}}}{r}\right)^3$ G (where $r$ is the radial distance from the WD) and estimated that a magnetic field strength of $\sim$875 G or $\sim1750$ G in the radio source region was necessary to produce cyclotron-maser emission. An electron density of $\geq10^{11}$ cm$^{-3}$ requires the higher magnetic field strength (see the discussion in Section 4 of \citealt{Benz1989}). TT Ari shows the same radio properties as EM Cyg, and the assumptions and estimates made in \citeauthor{Benz1989} are applicable to TT Ari.

\subsection{Source of the Emission}\label{sec:flarestars}

\begin{figure}
    \includegraphics[width=8cm]{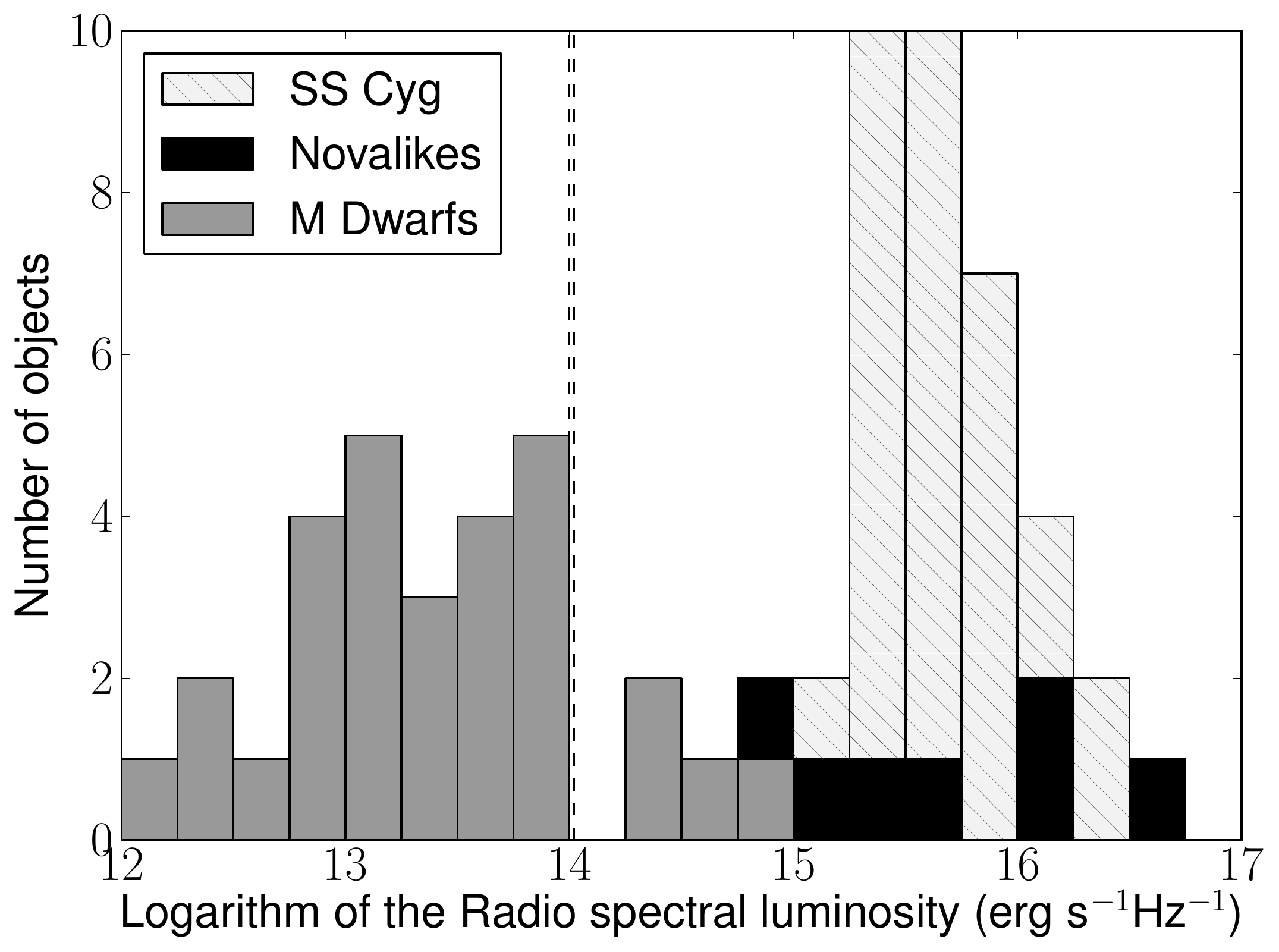}
    \caption{Histogram of the luminosities of all the flaring isolated M-dwarfs from \citet{McLean2012}, as well as the CVs observed in the radio since 2008. The dashed line at 10$^{14}\rm{\,erg\,s}^{-1}\rm{Hz}^{-1}$ indicates the upper edge of quiescence given by \citet{Gudel1993}.}
    \label{fig:KMdwarf_radio_fluxes}
\end{figure}

Besides the sources mentioned above, radio emission could also be produced by the secondary star. Flare stars are isolated dwarfs (including K- or M-type dwarfs) that produce radio flares believed to be caused by magnetic reconnections in the star's atmosphere (analogous to solar flares). As the secondary stars in CVs are K or M-type dwarfs, it is possible that the radio emission detected here could be from a flaring secondary -- particularly in the case of TT Ari. 

Figure \ref{fig:KMdwarf_radio_fluxes} shows a histogram of the peak radio luminosities of the isolated flaring M dwarfs from \citet{McLean2012} and all the high-sensitivity radio observations of non-magnetic CVs. The dashed line at 10$^{14}\mathrm{erg}\,\mathrm{s}^{-1}\,\mathrm{Hz}^{-1}$ indicates the upper-edge of the radio emission from quiescent M-dwarfs from \citet{Gudel1993}. As RW Sex and V603 Aql are not flaring in our observations and have luminosities that are significantly higher than the quiescent flare stars, we conclude that their radio emission is not produced by a flaring secondary. TT Ari, however, is clearly flaring and the variability time-scales, high brightness temperature and circular polarization fit the properties of flare stars -- particularly as they can produce flares that are up to 100\% circularly polarized (e.g. \citealt{Abada-Simon1997}). Although there is some overlap between the maxima of some of the flare stars and the novalike luminosities, TT Ari peaked at around 3.3$\times10^{16}\rm{\,erg\,s}^{-1}\rm{Hz}^{-1}$ which is 38 times higher than the brightest flare in \citet{McLean2012}. We think that it is thus unlikely that the radio emission of TT Ari is produced by a flaring secondary.

It should be noted that CV secondaries are tidally locked and have much higher rotation rates than isolated dwarfs, and that the impact of stellar rotation rates on magnetic fields is not well understood (see \citet{McLean2012} for a discussion). \citeauthor{McLean2012}, however, found that the radio emission of M0--M6 type dwarfs saturates at $10^{-7.5}L_{\rm{bolometric}}$ for rotation rates larger than $v\rm{sin}i\simeq5\,\mathrm{km\,s}^{-1}$, so the comparison between CV secondaries and isolated dwarfs is appropriate.

A final consideration is whether TT Ari has actually been misclassified as non-magnetic and the (stronger) magnetic fields play a more significant role in generating the radio emission. Magnetic CVs have shown variable and highly circularly polarized emission just like that of TT Ari. For example, \citet{Dulk1983} detected a 9.7mJy 100\% circularly-polarized radio flare at 4.9 GHz from the polar AM Her, which they concluded was probably due to a cyclotron maser. As mentioned previously however, a similar flare at 81\% CP was detected in the non-magnetic CV EM Cyg. Interestingly, the CV in our sample that is the most likely to be magnetic, V1084 Her, was the one source that was not detected. As CVs are not well studied in the radio, the radio properties of magnetic and non-magnetic CVs are not yet well defined and larger radio samples are needed. 

\section{Conclusion}\label{sec:conclusion}

We observed a sample of 4 novalikes at 6 GHz with the VLA and obtained a 75\% detection rate, which doubles the number of detections of non-magnetic CVs. These observations show that the sensitivity of previous radio surveys ($\sim1$ mJy) was too low to detect non-magnetic CVs and that many novalikes are in fact significant radio emitters.

V603 Aql, RW Sex and the old nova TT Ari were each detected in two epochs, as point sources, while V1084 Her was detected in neither of the two epochs. The distance uncertainty on V1084 Her is too large to tell if it is intrinsically faint or too far away.

The observations show a range of properties that suggest that more than one emission mechanism is responsible for the radio emission in our sample. In the literature, emission mechanisms that have been suggested for non-magnetic CVs include thermal emission, gyrosynchrotron and synchrotron emission, and cyclotron maser emission.

RW Sex was detected at approximately the same flux density in both epochs (33.6 $\mu\textrm{Jy\,beam}^{-1}$) and with a spectral index $\alpha=-0.5\pm$0.7 ($F=\nu^{\alpha}$). It is unlikely the emission is thermal emission, as the emitting region would need to be a factor 10$^2$--10$^3$ times the orbital separation to produce the observed brightness temperature. Gyrosynchrotron and cyclotron maser emission are consistent with our observations, so we cannot rule these emission mechanisms out. As RW Sex has a $3\sigma$ CP fraction upper-limit of 12.9\% and is not variable, however, we favour optically thin synchrotron emission.

V603 Aql was variable on timescales down to 217 s, with amplitudes of up to 61 $\mu\textrm{Jy\,beam}^{-1}$ and had a spectral index $\alpha=0.54\pm0.05$ in the first observation. In the second observation V603 Aql was not variable and the spectral index was flatter ($\alpha=0.16\pm0.08$). The emission is unlikely to be thermal emission by the same argument as for RW Sex, and based on causality arguments and the observed variability timescales it cannot be optically thick thermal emission. The $3\sigma$ upper-limit on the CP fraction is 12\% and the emission was variable, so we cannot rule out gyrosynchrotron or cyclotron maser emission. The radio detection is also consistent with optically thick synchrotron emission, which is consistent with the \citet{Koerding2008} prediction of a partially quenched, optically thick synchrotron jet. 

The two observations of TT Ari differed remarkably. The first showed a $\sim$10-minute flare with a peak flux density of 125.0$\pm16.2\,\mu\textrm{Jy\,beam}^{-1}$, that then declined to a 3$\sigma$ upper-limit of 49.5 $\mu\textrm{Jy\,beam}^{-1}$; $\sim$8 mins of the flare was circularly polarized (CP) and peaked at a CP fraction of 100\%. The flux in the second observation was higher (201.8--251.9 $\mu\textrm{Jy\,beam}^{-1}$) and the highest CP detection was 36.1$\pm10.0\,\mu\textrm{Jy\,beam}^{-1}$ (polarization fraction of 15\%). Radio behaviour like this has been seen in the magnetic CV AM Her \citep{Dulk1983,Chanmugam1987} and in another non-magnetic CV (EM Cyg; \citealt{Benz1989}).

The observed CP fraction for TT Ari is too high to be produced by synchrotron emission, but can be explained by either gyrosynchrotron or cyclotron maser emission. By the same arguments as for V603 Aql and the additional fact that thermal emission cannot produce CP, we can rule out thermal emission in TT Ari. However, as the properties of the two epochs suggest different emission mechanisms, and the spectrum of the second epoch is consistent with $\alpha$=2 to 6 GHz and $\alpha$=-0.1 at higher frequencies (which could be indicative of thermal emission with a turnover frequency at 6 GHz), we did consider the possibility that there is a thermal component to the emission in the second epoch. If this is the case, the observed spectrum is more consistent with a thin, dense shell than an extended, centrally concentrated stellar wind, and we would need to observe the evolution of the spectrum over multiple epochs to derive the mass of the emitting region. As an additional non-thermal component is necessary to produce the CP and observed variability, we favour pure gyrosynchrotron or cyclotron maser emission as emission mechanisms.

The high CP levels and variability shown by TT Ari are also consistent with flare star behaviour. For all three novalikes however, we conclude that although it is possible, the emission is unlikely to be produced in flares of the secondary star, as the luminosities are significantly higher than those seen in both flaring and quiescent flare stars.

We did not find a radio-optical flux relation or a radio flux-distance relation, but this may change as the sample of radio detections of non-magnetic CVs increases.

As we have demonstrated, it is now possible to detect non-magnetic CVs with the VLA. Further observations of CVs will help establish the nature of the radio emission, which could then be used to study accretion and possibly outflow physics in these nearby, numerous and non-relativistic compact accretors.

\section*{Acknowledgements}

We thank the referee for their prompt and positive response. Thank you also to Cameron Van Eck for many helpful discussions.

This work is part of the research programme NWO VIDI grant Nr. 2013/15390/EW, which is financed by the Netherlands Organisation for Scientific Research (NWO, Nederlandse Organisatie voor Wetenschappelijk Onderzoek).

DLC gratefully acknowledges funding from the Erasmus Mundus Programme SAPIENT. JCAMJ is the recipient of an Australian Research Council Future Fellowship (FT140101082), and also acknowledges support from an Australian Research Council Discovery Grant (DP120102393). GRS acknowledges support from an NSERC Discovery Grant.

The National Radio Astronomy Observatory is a facility of the National Science Foundation operated under cooperative agreement by Associated Universities, Inc. This research has made use of NASA's Astrophysics Data System Bibliographic Services, as well as the SIMBAD database, operated at CDS, Strasbourg, France \citep{Wenger2000}.

\bibliographystyle{mn2e.bst}
\bibliography{novalikes}

\begin{thebibliography}{}

\bibitem[\protect\citeauthoryear{{Abada-Simon} \& {Aubier}}{{Abada-Simon} \&
  {Aubier}}{1997}]{Abada-Simon1997}
{Abada-Simon} M.,  {Aubier} M.,  1997, \aaps, 125, 511

\bibitem[\protect\citeauthoryear{{Abada-Simon}, {Lecacheux}, {Bastian},
  {Bookbinder} \& {Dulk}}{{Abada-Simon} et~al.}{1993}]{Abada-Simon1993}
{Abada-Simon} M.,  {Lecacheux} A.,  {Bastian} T.~S.,  {Bookbinder} J.~A.,
  {Dulk} G.~A.,  1993, \apj, 406, 692

\bibitem[\protect\citeauthoryear{{Ak}, {Bilir}, {Ak} \& {Eker}}{{Ak}
  et~al.}{2008}]{Ak2008}
{Ak} T.,  {Bilir} S.,  {Ak} S.,    {Eker} Z.,  2008, \na, 13, 133

\bibitem[\protect\citeauthoryear{{Andronov}, {Ostrova}, {Kim} \&
  {Burwitz}}{{Andronov} et~al.}{2005}]{Andronov2005}
{Andronov} I.~L.,  {Ostrova} N.~I.,  {Kim} Y.-G.,    {Burwitz} V.,  2005,
  Journal of Astronomy and Space Sciences, 22, 211

\bibitem[\protect\citeauthoryear{{Arenas}, {Catal{\'a}n}, {Augusteijn} \&
  {Retter}}{{Arenas} et~al.}{2000}]{Arenas2000}
{Arenas} J.,  {Catal{\'a}n} M.~S.,  {Augusteijn} T.,    {Retter} A.,  2000,
  \mnras, 311, 135

\bibitem[\protect\citeauthoryear{{Baskill}, {Wheatley} \& {Osborne}}{{Baskill}
  et~al.}{2005}]{Baskill2005}
{Baskill} D.~S.,  {Wheatley} P.~J.,    {Osborne} J.~P.,  2005, \mnras, 357, 626

\bibitem[\protect\citeauthoryear{{Benz}, {Fuerst} \& {Kiplinger}}{{Benz}
  et~al.}{1983}]{Benz1983}
{Benz} A.~O.,  {Fuerst} E.,    {Kiplinger} A.~L.,  1983, \nat, 302, 45

\bibitem[\protect\citeauthoryear{{Benz}, {Gudel} \& {Mattei}}{{Benz}
  et~al.}{1996}]{Benz1996}
{Benz} A.~O.,  {Gudel} M.,    {Mattei} J.~A.,  1996, in {Taylor} A.~R.,
  {Paredes} J.~M.,  eds, Radio Emission from the Stars and the Sun Vol.~93 of
  Astronomical Society of the Pacific Conference Series, {Radio Emission of
  Dwarf Novae}.
p.~188

\bibitem[\protect\citeauthoryear{{Benz} \& {Guedel}}{{Benz} \&
  {Guedel}}{1989}]{Benz1989}
{Benz} A.~O.,  {Guedel} M.,  1989, \aap, 218, 137

\bibitem[\protect\citeauthoryear{{Beuermann}, {Stasiewski} \&
  {Schwope}}{{Beuermann} et~al.}{1992}]{Beuermann1992}
{Beuermann} K.,  {Stasiewski} U.,    {Schwope} A.~D.,  1992, \aap, 256, 433

\bibitem[\protect\citeauthoryear{{Borczyk}, {Schwarzenberg-Czerny} \&
  {Szkody}}{{Borczyk} et~al.}{2003}]{Borczyk2003}
{Borczyk} W.,  {Schwarzenberg-Czerny} A.,    {Szkody} P.,  2003, \aap, 405, 663

\bibitem[\protect\citeauthoryear{{Chanmugam}}{{Chanmugam}}{1987}]{Chanmugam198%
7}
{Chanmugam} G.,  1987, \apss, 130, 53

\bibitem[\protect\citeauthoryear{{Corbel}, {Fender}, {Tzioumis}, {Nowak},
  {McIntyre}, {Durouchoux} \& {Sood}}{{Corbel} et~al.}{2000}]{Corbel2000}
{Corbel} S.,  {Fender} R.~P.,  {Tzioumis} A.~K.,  {Nowak} M.,  {McIntyre} V.,
  {Durouchoux} P.,    {Sood} R.,  2000, \aap, 359, 251

\bibitem[\protect\citeauthoryear{{Cordova}, {Hjellming} \& {Mason}}{{Cordova}
  et~al.}{1983}]{Cordova1983}
{Cordova} F.~A.,  {Hjellming} R.~M.,    {Mason} K.~O.,  1983, \pasp, 95, 69

\bibitem[\protect\citeauthoryear{{Cowley}, {Crampton}, {Hutchings} \&
  {Marlborough}}{{Cowley} et~al.}{1975}]{Cowley1975}
{Cowley} A.~P.,  {Crampton} D.,  {Hutchings} J.~B.,    {Marlborough} J.~M.,
  1975, \apj, 195, 413

\bibitem[\protect\citeauthoryear{{Cutri}, {Skrutskie}, {van Dyk}, {Beichman},
  {Carpenter}, {Chester}, {Cambresy}, {Evans}, {Fowler}, {Gizis}, {Howard},
  {Huchra}, {Jarrett}, {Kopan}, {Kirkpatrick}, {Light}, {Marsh} \&
  {McCallon}}{{Cutri} et~al.}{2003}]{Cutri2003}
{Cutri} R.~M.,  {Skrutskie} M.~F.,  {van Dyk} S.,  {Beichman} C.~A.,
  {Carpenter} J.~M.,  {Chester} T.,  {Cambresy} L.,  {Evans} T.,  {Fowler} J.,
  {Gizis} J.,  {Howard} E.,  {Huchra} J.,  {Jarrett} T.,  {Kopan} E.~L.,
  {Kirkpatrick} J.~D.,  {Light} R.~M.,  {Marsh} K.~A.,    {McCallon} H.,  2003,
  VizieR Online Data Catalog, 2246, 0

\bibitem[\protect\citeauthoryear{{de Martino}, {Sala}, {Balman}, {Bernardini},
  {Bianchini}, {Bode}, {Bonnet-Bidaud}, {Falanga}, {Greiner}, {Groot},
  {Hernanz}, {Israel}, {Jose}, {Motch}, {Mouchet}, {Norton} \& {Nucita}}{{de
  Martino} et~al.}{2015}]{Martino2015}
{de Martino} D.,  {Sala} G.,  {Balman} S.,  {Bernardini} F.,  {Bianchini} A.,
  {Bode} M.,  {Bonnet-Bidaud} J.-M.,  {Falanga} M.,  {Greiner} J.,  {Groot} P.,
   {Hernanz} M.,  {Israel} G.,  {Jose} J.,  {Motch} C.,  {Mouchet} M.,
  {Norton} A.~J.,    {Nucita} A.,  2015, ArXiv e-prints

\bibitem[\protect\citeauthoryear{{Dhillon}, {Smith} \& {Marsh}}{{Dhillon}
  et~al.}{2013}]{Dhillon2013}
{Dhillon} V.~S.,  {Smith} D.~A.,    {Marsh} T.~R.,  2013, \mnras, 428, 3559

\bibitem[\protect\citeauthoryear{{Dulk}, {Bastian} \& {Chanmugam}}{{Dulk}
  et~al.}{1983}]{Dulk1983}
{Dulk} G.~A.,  {Bastian} T.~S.,    {Chanmugam} G.,  1983, \apj, 273, 249

\bibitem[\protect\citeauthoryear{{Echevarria}}{{Echevarria}}{1987}]{Echevarria%
1987}
{Echevarria} J.,  1987, \apss, 130, 103

\bibitem[\protect\citeauthoryear{{Falcke}, {K{\"o}rding} \& {Markoff}}{{Falcke}
  et~al.}{2004}]{Falcke2004}
{Falcke} H.,  {K{\"o}rding} E.,    {Markoff} S.,  2004, \aap, 414, 895

\bibitem[\protect\citeauthoryear{{Fender}}{{Fender}}{2001}]{Fender2001}
{Fender} R.~P.,  2001, \mnras, 322, 31

\bibitem[\protect\citeauthoryear{{Fender}, {Belloni} \& {Gallo}}{{Fender}
  et~al.}{2004}]{Fender2004}
{Fender} R.~P.,  {Belloni} T.~M.,    {Gallo} E.,  2004, \mnras, 355, 1105

\bibitem[\protect\citeauthoryear{{Fuerst}, {Benz}, {Hirth}, {Kiplinger} \&
  {Geffert}}{{Fuerst} et~al.}{1986}]{Fuerst1986}
{Fuerst} E.,  {Benz} A.,  {Hirth} W.,  {Kiplinger} A.,    {Geffert} M.,  1986,
  \aap, 154, 377

\bibitem[\protect\citeauthoryear{{G{\"a}nsicke}, {Sion}, {Beuermann}, {Fabian},
  {Cheng} \& {Krautter}}{{G{\"a}nsicke} et~al.}{1999}]{Gaensicke1999}
{G{\"a}nsicke} B.~T.,  {Sion} E.~M.,  {Beuermann} K.,  {Fabian} D.,  {Cheng}
  F.~H.,    {Krautter} J.,  1999, \aap, 347, 178

\bibitem[\protect\citeauthoryear{{Gnedin}, {Borisov} \&
  {Natsvlishvili}}{{Gnedin} et~al.}{1990}]{Gnedin1990}
{Gnedin} Y.~N.,  {Borisov} N.~V.,    {Natsvlishvili} T.~M.,  1990, Soviet
  Astronomy Letters, 16, 272

\bibitem[\protect\citeauthoryear{{Greenstein} \& {Oke}}{{Greenstein} \&
  {Oke}}{1982}]{Greenstein1982}
{Greenstein} J.~L.,  {Oke} J.~B.,  1982, \apj, 258, 209

\bibitem[\protect\citeauthoryear{{Gudel}, {Schmitt}, {Bookbinder} \&
  {Fleming}}{{Gudel} et~al.}{1993}]{Gudel1993}
{Gudel} M.,  {Schmitt} J.~H.~M.~M.,  {Bookbinder} J.~A.,    {Fleming} T.~A.,
  1993, \apj, 415, 236

\bibitem[\protect\citeauthoryear{{H{\= o}shi}}{{H{\= o}shi}}{1979}]{Hoshi1979}
{H{\= o}shi} R.,  1979, Progress of Theoretical Physics, 61, 1307

\bibitem[\protect\citeauthoryear{{Haefner} \& {Metz}}{{Haefner} \&
  {Metz}}{1985}]{Haefner1985}
{Haefner} R.,  {Metz} K.,  1985, \aap, 145, 311

\bibitem[\protect\citeauthoryear{{Han} \& {Hjellming}}{{Han} \&
  {Hjellming}}{1992}]{Han1992}
{Han} X.,  {Hjellming} R.~M.,  1992, \apj, 400, 304

\bibitem[\protect\citeauthoryear{{Harrison}, {Bornak}, {McArthur} \&
  {Benedict}}{{Harrison} et~al.}{2013}]{Harrison2013}
{Harrison} T.~E.,  {Bornak} J.,  {McArthur} B.~E.,    {Benedict} G.~F.,  2013,
  \apj, 767, 7

\bibitem[\protect\citeauthoryear{{Hartley}, {Drew}, {Long}, {Knigge} \&
  {Proga}}{{Hartley} et~al.}{2002}]{Hartley2002}
{Hartley} L.~E.,  {Drew} J.~E.,  {Long} K.~S.,  {Knigge} C.,    {Proga} D.,
  2002, \mnras, 332, 127

\bibitem[\protect\citeauthoryear{{H{\o}g}, {Fabricius}, {Makarov}, {Urban},
  {Corbin}, {Wycoff}, {Bastian}, {Schwekendiek} \& {Wicenec}}{{H{\o}g}
  et~al.}{2000}]{Hog2000}
{H{\o}g} E.,  {Fabricius} C.,  {Makarov} V.~V.,  {Urban} S.,  {Corbin} T.,
  {Wycoff} G.,  {Bastian} U.,  {Schwekendiek} P.,    {Wicenec} A.,  2000, \aap,
  355, L27

\bibitem[\protect\citeauthoryear{{Hutchings} \& {Cowley}}{{Hutchings} \&
  {Cowley}}{2007}]{Hutchings2007}
{Hutchings} J.~B.,  {Cowley} A.~P.,  2007, \aj, 133, 1204

\bibitem[\protect\citeauthoryear{{Johnson}, {Schaefer}, {Kroll} \&
  {Henden}}{{Johnson} et~al.}{2014}]{Johnson2014}
{Johnson} C.~B.,  {Schaefer} B.~E.,  {Kroll} P.,    {Henden} A.~A.,  2014,
  \apjl, 780, L25

\bibitem[\protect\citeauthoryear{{Kafka}, {Hoard}, {Honeycutt} \&
  {Deliyannis}}{{Kafka} et~al.}{2009}]{Kafka2009}
{Kafka} S.,  {Hoard} D.~W.,  {Honeycutt} R.~K.,    {Deliyannis} C.~P.,  2009,
  \aj, 137, 197

\bibitem[\protect\citeauthoryear{{K{\"o}rding}}{{K{\"o}rding}}{2008}]{Koerding%
2008b}
{K{\"o}rding} E.,  2008, in Microquasars and Beyond {Connections from
  supermassive black holes to white dwarfs}.
p.~23

\bibitem[\protect\citeauthoryear{{K{\"o}rding}, {Rupen}, {Knigge}, {Fender},
  {Dhawan}, {Templeton} \& {Muxlow}}{{K{\"o}rding} et~al.}{2008}]{Koerding2008}
{K{\"o}rding} E.,  {Rupen} M.,  {Knigge} C.,  {Fender} R.,  {Dhawan} V.,
  {Templeton} M.,    {Muxlow} T.,  2008, Science, 320, 1318

\bibitem[\protect\citeauthoryear{{K{\"o}rding}, {Fender} \&
  {Migliari}}{{K{\"o}rding} et~al.}{2006}]{Koerding2006}
{K{\"o}rding} E.~G.,  {Fender} R.~P.,    {Migliari} S.,  2006, \mnras, 369,
  1451

\bibitem[\protect\citeauthoryear{{K{\"o}rding}, {Knigge}, {Tzioumis} \&
  {Fender}}{{K{\"o}rding} et~al.}{2011}]{Koerding2011}
{K{\"o}rding} E.~G.,  {Knigge} C.,  {Tzioumis} T.,    {Fender} R.,  2011,
  \mnras, 418, L129

\bibitem[\protect\citeauthoryear{{K{\"o}rding}, {Migliari}, {Fender},
  {Belloni}, {Knigge} \& {McHardy}}{{K{\"o}rding} et~al.}{2007}]{Koerding2007}
{K{\"o}rding} E.~G.,  {Migliari} S.,  {Fender} R.,  {Belloni} T.,  {Knigge} C.,
     {McHardy} I.,  2007, \mnras, 380, 301

\bibitem[\protect\citeauthoryear{{Lasota}}{{Lasota}}{2001}]{Lasota2001}
{Lasota} J.-P.,  2001, \nar, 45, 449

\bibitem[\protect\citeauthoryear{{Linnell}, {Godon}, {Hubeny}, {Sion} \&
  {Szkody}}{{Linnell} et~al.}{2010}]{Linnell2010}
{Linnell} A.~P.,  {Godon} P.,  {Hubeny} I.,  {Sion} E.~M.,    {Szkody} P.,
  2010, \apj, 719, 271

\bibitem[\protect\citeauthoryear{{Livio}}{{Livio}}{1999}]{Livio1999}
{Livio} M.,  1999, \physrep, 311, 225

\bibitem[\protect\citeauthoryear{{Longair}}{{Longair}}{2011}]{Longair2011}
{Longair} M.~S.,  2011, {High Energy Astrophysics}

\bibitem[\protect\citeauthoryear{{Mason} \& {Gray}}{{Mason} \&
  {Gray}}{2007}]{Mason2007}
{Mason} P.~A.,  {Gray} C.~L.,  2007, \apj, 660, 662

\bibitem[\protect\citeauthoryear{{McLean}, {Berger} \& {Reiners}}{{McLean}
  et~al.}{2012}]{McLean2012}
{McLean} M.,  {Berger} E.,    {Reiners} A.,  2012, \apj, 746, 23

\bibitem[\protect\citeauthoryear{{McMullin}, {Waters}, {Schiebel}, {Young} \&
  {Golap}}{{McMullin} et~al.}{2007}]{McMullin2007}
{McMullin} J.~P.,  {Waters} B.,  {Schiebel} D.,  {Young} W.,    {Golap} K.,
  2007, in {Shaw} R.~A.,  {Hill} F.,   {Bell} D.~J.,  eds, Astronomical Data
  Analysis Software and Systems XVI Vol.~376 of Astronomical Society of the
  Pacific Conference Series, {CASA Architecture and Applications}.
p.~127

\bibitem[\protect\citeauthoryear{{Meintjes} \& {Venter}}{{Meintjes} \&
  {Venter}}{2005}]{Meintjes2005}
{Meintjes} P.~J.,  {Venter} L.~A.,  2005, \mnras, 360, 573

\bibitem[\protect\citeauthoryear{{Merloni}, {Heinz} \& {di Matteo}}{{Merloni}
  et~al.}{2003}]{Merloni2003}
{Merloni} A.,  {Heinz} S.,    {di Matteo} T.,  2003, \mnras, 345, 1057

\bibitem[\protect\citeauthoryear{{Mickaelian}, {Balayan}, {Ilovaisky},
  {Chevalier}, {V{\'e}ron-Cetty} \& {V{\'e}ron}}{{Mickaelian}
  et~al.}{2002}]{Mickaelian2002}
{Mickaelian} A.~M.,  {Balayan} S.~K.,  {Ilovaisky} S.~A.,  {Chevalier} C.,
  {V{\'e}ron-Cetty} M.-P.,    {V{\'e}ron} P.,  2002, \aap, 381, 894

\bibitem[\protect\citeauthoryear{{Migliari} \& {Fender}}{{Migliari} \&
  {Fender}}{2006}]{Migliari2006}
{Migliari} S.,  {Fender} R.~P.,  2006, \mnras, 366, 79

\bibitem[\protect\citeauthoryear{{Miller-Jones}, {Sivakoff}, {Altamirano},
  {K{\"o}rding}, {Krimm}, {Maitra}, {Remillard}, {Russell}, {Tudose}, {Dhawan},
  {Fender}, {Heinz}, {Markoff}, {Migliari}, {Rupen} \&
  {Sarazin}}{{Miller-Jones} et~al.}{2011}]{Miller-Jones2011}
{Miller-Jones} J.~C.~A.,  {Sivakoff} G.~R.,  {Altamirano} D.,  {K{\"o}rding}
  E.~G.,  {Krimm} H.~A.,  {Maitra} D.,  {Remillard} R.~A.,  {Russell} D.~M.,
  {Tudose} V.,  {Dhawan} V.,  {Fender} R.~P.,  {Heinz} S.,  {Markoff} S.,
  {Migliari} S.,  {Rupen} M.~P.,    {Sarazin} C.~L.,  2011, in {Romero} G.~E.,
  {Sunyaev} R.~A.,   {Belloni} T.,  eds, IAU Symposium Vol.~275 of IAU
  Symposium, {Investigating accretion disk - radio jet coupling across the
  stellar mass scale}.
pp 224--232

\bibitem[\protect\citeauthoryear{{Miller-Jones}, {Sivakoff}, {Knigge},
  {K{\"o}rding}, {Templeton} \& {Waagen}}{{Miller-Jones}
  et~al.}{2013}]{Miller-Jones2013}
{Miller-Jones} J.~C.~A.,  {Sivakoff} G.~R.,  {Knigge} C.,  {K{\"o}rding} E.~G.,
   {Templeton} M.,    {Waagen} E.~O.,  2013, Science, 340, 950

\bibitem[\protect\citeauthoryear{{Mukai} \& {Orio}}{{Mukai} \&
  {Orio}}{2005}]{Mukai2005}
{Mukai} K.,  {Orio} M.,  2005, \apj, 622, 602

\bibitem[\protect\citeauthoryear{{Naylor}, {Koch-Miramond}, {Ringwald} \&
  {Evans}}{{Naylor} et~al.}{1996}]{Naylor1996}
{Naylor} T.,  {Koch-Miramond} L.,  {Ringwald} F.~A.,    {Evans} A.,  1996,
  \mnras, 282, 873

\bibitem[\protect\citeauthoryear{{Nelson} \& {Spencer}}{{Nelson} \&
  {Spencer}}{1988}]{Nelson1988}
{Nelson} R.~F.,  {Spencer} R.~E.,  1988, \mnras, 234, 1105

\bibitem[\protect\citeauthoryear{{Osaki}}{{Osaki}}{1974}]{Osaki1974}
{Osaki} Y.,  1974, \pasj, 26, 429

\bibitem[\protect\citeauthoryear{{Patterson}, {Fenton}, {Thorstensen},
  {Harvey}, {Skillman}, {Fried}, {Monard}, {O'Donoghue}, {Beshore}, {Martin},
  {Niarchos}, {Vanmunster}, {Foote}, {Bolt}, {Rea}, {Cook}, {Butterworth} \&
  {Wood}}{{Patterson} et~al.}{2002}]{Patterson2002}
{Patterson} J.,  {Fenton} W.~H.,  {Thorstensen} J.~R.,  {Harvey} D.~A.,
  {Skillman} D.~R.,  {Fried} R.~E.,  {Monard} B.,  {O'Donoghue} D.,  {Beshore}
  E.,  {Martin} B.,  {Niarchos} P.,  {Vanmunster} T.,  {Foote} J.,  {Bolt} G.,
  {Rea} R.,  {Cook} L.~M.,  {Butterworth} N.,    {Wood} M.,  2002, \pasp, 114,
  1364

\bibitem[\protect\citeauthoryear{{Patterson}, {Kemp}, {Saad}, {Skillman},
  {Harvey}, {Fried}, {Thorstensen} \& {Ashley}}{{Patterson}
  et~al.}{1997}]{Patterson1997}
{Patterson} J.,  {Kemp} J.,  {Saad} J.,  {Skillman} D.~R.,  {Harvey} D.,
  {Fried} R.,  {Thorstensen} J.~R.,    {Ashley} R.,  1997, \pasp, 109, 468

\bibitem[\protect\citeauthoryear{{Patterson} \& {Richman}}{{Patterson} \&
  {Richman}}{1991}]{Patterson1991}
{Patterson} J.,  {Richman} H.,  1991, \pasp, 103, 735

\bibitem[\protect\citeauthoryear{{Patterson}, {Thomas}, {Skillman} \&
  {Diaz}}{{Patterson} et~al.}{1993}]{Patterson1993}
{Patterson} J.,  {Thomas} G.,  {Skillman} D.~R.,    {Diaz} M.,  1993, \apjs,
  86, 235

\bibitem[\protect\citeauthoryear{{Patterson}, {Uthas}, {Kemp}, {de Miguel},
  {Krajci}, {Foote}, {Hambsch} \& {Campbell}}{{Patterson}
  et~al.}{2013}]{Patterson2013}
{Patterson} J.,  {Uthas} H.,  {Kemp} J.,  {de Miguel} E.,  {Krajci} T.,
  {Foote} J.,  {Hambsch} F.-J.,    {Campbell} 2013, \mnras, 434, 1902

\bibitem[\protect\citeauthoryear{{Pavelin}, {Spencer} \& {Davis}}{{Pavelin}
  et~al.}{1994}]{Pavelin1994}
{Pavelin} P.~E.,  {Spencer} R.~E.,    {Davis} R.~J.,  1994, \mnras, 269, 779

\bibitem[\protect\citeauthoryear{{Prinja}, {Knigge}, {Ringwald} \&
  {Wade}}{{Prinja} et~al.}{2000}]{Prinja2000}
{Prinja} R.~K.,  {Knigge} C.,  {Ringwald} F.~A.,    {Wade} R.~A.,  2000,
  \mnras, 318, 368

\bibitem[\protect\citeauthoryear{{Prinja}, {Long}, {Froning}, {Knigge},
  {Witherick}, {Clark} \& {Ringwald}}{{Prinja} et~al.}{2003}]{Prinja2003}
{Prinja} R.~K.,  {Long} K.~S.,  {Froning} C.~S.,  {Knigge} C.,  {Witherick}
  D.~K.,  {Clark} J.~S.,    {Ringwald} F.~A.,  2003, \mnras, 340, 551

\bibitem[\protect\citeauthoryear{{Prinja} \& {Rosen}}{{Prinja} \&
  {Rosen}}{1995}]{Prinja1995}
{Prinja} R.~K.,  {Rosen} R.,  1995, \mnras, 273, 461

\bibitem[\protect\citeauthoryear{{Retter} \& {Naylor}}{{Retter} \&
  {Naylor}}{2000}]{Retter2000}
{Retter} A.,  {Naylor} T.,  2000, \mnras, 319, 510

\bibitem[\protect\citeauthoryear{{Ritter} \& {Kolb}}{{Ritter} \&
  {Kolb}}{2003}]{Ritter2003}
{Ritter} H.,  {Kolb} U.,  2003, \aap, 404, 301

\bibitem[\protect\citeauthoryear{{Robinson} \& {Cordova}}{{Robinson} \&
  {Cordova}}{1994}]{Robinson1994}
{Robinson} C.~R.,  {Cordova} F.~A.,  1994, in {Shafter} A.~W.,  ed.,
  Interacting Binary Stars Vol.~56 of Astronomical Society of the Pacific
  Conference Series, {The X-ray Behavior of Two CVs: TT Ari and DP Leo}.
p.~146

\bibitem[\protect\citeauthoryear{{Rodr{\'{\i}}guez-Gil}, {Casares},
  {Mart{\'{\i}}nez-Pais}, {Hakala} \& {Steeghs}}{{Rodr{\'{\i}}guez-Gil}
  et~al.}{2001}]{Rodriguez-Gil2001}
{Rodr{\'{\i}}guez-Gil} P.,  {Casares} J.,  {Mart{\'{\i}}nez-Pais} I.~G.,
  {Hakala} P.,    {Steeghs} D.,  2001, \apjl, 548, L49

\bibitem[\protect\citeauthoryear{{Rodr{\'{\i}}guez-Gil}, {G{\"a}nsicke},
  {Hagen}, {Araujo-Betancor}, {Aungwerojwit}, {Allende Prieto}, {Boyd},
  {Casares}, {Engels}, {Giannakis} \& {Harlaftis}}{{Rodr{\'{\i}}guez-Gil}
  et~al.}{2007}]{Rodriguez-Gil2007}
{Rodr{\'{\i}}guez-Gil} P.,  {G{\"a}nsicke} B.~T.,  {Hagen} H.-J.,
  {Araujo-Betancor} S.,  {Aungwerojwit} A.,  {Allende Prieto} C.,  {Boyd} D.,
  {Casares} J.,  {Engels} D.,  {Giannakis} O.,    {Harlaftis} E.~T.,  2007,
  \mnras, 377, 1747

\bibitem[\protect\citeauthoryear{{Rodr{\'{\i}}guez-Gil} \&
  {Mart{\'{\i}}nez-Pais}}{{Rodr{\'{\i}}guez-Gil} \&
  {Mart{\'{\i}}nez-Pais}}{2002}]{Rodriguez-Gil2002}
{Rodr{\'{\i}}guez-Gil} P.,  {Mart{\'{\i}}nez-Pais} I.~G.,  2002, \mnras, 337,
  209

\bibitem[\protect\citeauthoryear{{Rodr{\'{\i}}guez-Gil}, {Mart{\'{\i}}nez-Pais}
  \& {de la Cruz Rodr{\'{\i}}guez}}{{Rodr{\'{\i}}guez-Gil}
  et~al.}{2009}]{Rodriguez-Gil2009}
{Rodr{\'{\i}}guez-Gil} P.,  {Mart{\'{\i}}nez-Pais} I.~G.,    {de la Cruz
  Rodr{\'{\i}}guez} J.,  2009, \mnras, 395, 973

\bibitem[\protect\citeauthoryear{{Russell}, {Miller-Jones}, {Curran}, {Soria},
  {Altamirano}, {Corbel}, {Coriat}, {Moin}, {Russell}, {Sivakoff},
  {Slaven-Blair}, {Belloni}, {Fender}, {Heinz}, {Jonker}, {Krimm} \&
  {K{\"o}rding}}{{Russell} et~al.}{2015}]{Russell2015}
{Russell} T.~D.,  {Miller-Jones} J.~C.~A.,  {Curran} P.~A.,  {Soria} R.,
  {Altamirano} D.,  {Corbel} S.,  {Coriat} M.,  {Moin} A.,  {Russell} D.~M.,
  {Sivakoff} G.~R.,  {Slaven-Blair} T.~J.,  {Belloni} T.~M.,  {Fender} R.~P.,
  {Heinz} S.,  {Jonker} P.~G.,  {Krimm} H.~A.,    {K{\"o}rding} E.~G.,  2015,
  \mnras, 450, 1745

\bibitem[\protect\citeauthoryear{{Shafter}, {Szkody}, {Liebert}, {Penning},
  {Bond} \& {Grauer}}{{Shafter} et~al.}{1985}]{Shafter1985}
{Shafter} A.~W.,  {Szkody} P.,  {Liebert} J.,  {Penning} W.~R.,  {Bond} H.~E.,
    {Grauer} A.~D.,  1985, \apj, 290, 707

\bibitem[\protect\citeauthoryear{{Shara}, {Livio}, {Moffat} \& {Orio}}{{Shara}
  et~al.}{1986}]{Shara1986}
{Shara} M.~M.,  {Livio} M.,  {Moffat} A.~F.~J.,    {Orio} M.,  1986, \apj, 311,
  163

\bibitem[\protect\citeauthoryear{{Smak}}{{Smak}}{1971}]{Smak1971}
{Smak} J.,  1971, \acta, 21, 15

\bibitem[\protect\citeauthoryear{{Soker} \& {Lasota}}{{Soker} \&
  {Lasota}}{2004}]{Soker2004}
{Soker} N.,  {Lasota} J.-P.,  2004, \aap, 422, 1039

\bibitem[\protect\citeauthoryear{{Strohmeier}, {Kippenhahn} \&
  {Geyer}}{{Strohmeier} et~al.}{1957}]{Strohmeier1957}
{Strohmeier} W.,  {Kippenhahn} R.,    {Geyer} E.,  1957, K1. Ver\"{o}ff.
  Sternwarte Bamberg Nr. 18

\bibitem[\protect\citeauthoryear{{Strope}, {Schaefer} \& {Henden}}{{Strope}
  et~al.}{2010}]{Strope2010}
{Strope} R.~J.,  {Schaefer} B.~E.,    {Henden} A.~A.,  2010, \aj, 140, 34

\bibitem[\protect\citeauthoryear{{Tremko}, {Andronov}, {Luthardt}, {Pajdosz},
  {Patkos}, {Roessiger} \& {Zola}}{{Tremko} et~al.}{1992}]{Tremko1992}
{Tremko} J.,  {Andronov} I.~L.,  {Luthardt} R.,  {Pajdosz} G.,  {Patkos} L.,
  {Roessiger} S.,    {Zola} S.,  1992, Information Bulletin on Variable Stars,
  3763, 1

\bibitem[\protect\citeauthoryear{{Turner}}{{Turner}}{1985}]{Turner1985}
{Turner} K.~C.,  1985, in {Hjellming} R.~M.,  {Gibson} D.~M.,  eds, Radio Stars
  Vol.~116 of Astrophysics and Space Science Library, {12 cm Observations of
  Stellar Radio Sources}.
p.~283

\bibitem[\protect\citeauthoryear{{Udalski} \& {Schwarzenberg-Czerny}}{{Udalski}
  \& {Schwarzenberg-Czerny}}{1989}]{Udalski1989}
{Udalski} A.,  {Schwarzenberg-Czerny} A.,  1989, \acta, 39, 125

\bibitem[\protect\citeauthoryear{{van Leeuwen}}{{van
  Leeuwen}}{2007}]{vanleeuwen2007}
{van Leeuwen} F.,  2007, \aap, 474, 653

\bibitem[\protect\citeauthoryear{{van Teeseling}, {Beuermann} \&
  {Verbunt}}{{van Teeseling} et~al.}{1996}]{vanTeeseling1996}
{van Teeseling} A.,  {Beuermann} K.,    {Verbunt} F.,  1996, \aap, 315, 467

\bibitem[\protect\citeauthoryear{{Vitello} \& {Shlosman}}{{Vitello} \&
  {Shlosman}}{1993}]{Vitello1993}
{Vitello} P.,  {Shlosman} I.,  1993, \apj, 410, 815

\bibitem[\protect\citeauthoryear{{Vogt}, {Chen{\'e}}, {Moffat}, {Matthews},
  {Kuschnig}, {Guenther}, {Rowe}, {Rucinski}, {Sasselov} \& {Weiss}}{{Vogt}
  et~al.}{2013}]{Vogt2013}
{Vogt} N.,  {Chen{\'e}} A.-N.,  {Moffat} A.~F.~J.,  {Matthews} J.~M.,
  {Kuschnig} R.,  {Guenther} D.~B.,  {Rowe} J.~F.,  {Rucinski} S.~M.,
  {Sasselov} D.,    {Weiss} W.~W.,  2013, Astronomische Nachrichten, 334, 1101

\bibitem[\protect\citeauthoryear{{Warner}}{{Warner}}{1995}]{Warner1995}
{Warner} B.,  1995, Cambridge Astrophysics Series, 28

\bibitem[\protect\citeauthoryear{{Wenger}, {Ochsenbein}, {Egret}, {Dubois},
  {Bonnarel}, {Borde}, {Genova}, {Jasniewicz}, {Lalo{\"e}}, {Lesteven} \&
  {Monier}}{{Wenger} et~al.}{2000}]{Wenger2000}
{Wenger} M.,  {Ochsenbein} F.,  {Egret} D.,  {Dubois} P.,  {Bonnarel} F.,
  {Borde} S.,  {Genova} F.,  {Jasniewicz} G.,  {Lalo{\"e}} S.,  {Lesteven} S.,
    {Monier} R.,  2000, \aaps, 143, 9

\bibitem[\protect\citeauthoryear{{Wright}, {Stewart}, {Nelson}, {Slee} \&
  {Cropper}}{{Wright} et~al.}{1988}]{Wright1988}
{Wright} A.~E.,  {Stewart} R.~T.,  {Nelson} G.~J.,  {Slee} O.~B.,    {Cropper}
  M.,  1988, \mnras, 231, 319

\bibitem[\protect\citeauthoryear{{Wu}, {Li}, {Ding}, {Zhang} \& {Li}}{{Wu}
  et~al.}{2002}]{Wu2002}
{Wu} X.,  {Li} Z.,  {Ding} Y.,  {Zhang} Z.,    {Li} Z.,  2002, \apj, 569, 418

\end{thebibliography}

\label{lastpage}

\appendix

\section{Radio maps}

\begin{figure*}
  \subfloat{
  \begin{minipage}{70mm}
    \centering
    \includegraphics[width=\textwidth]{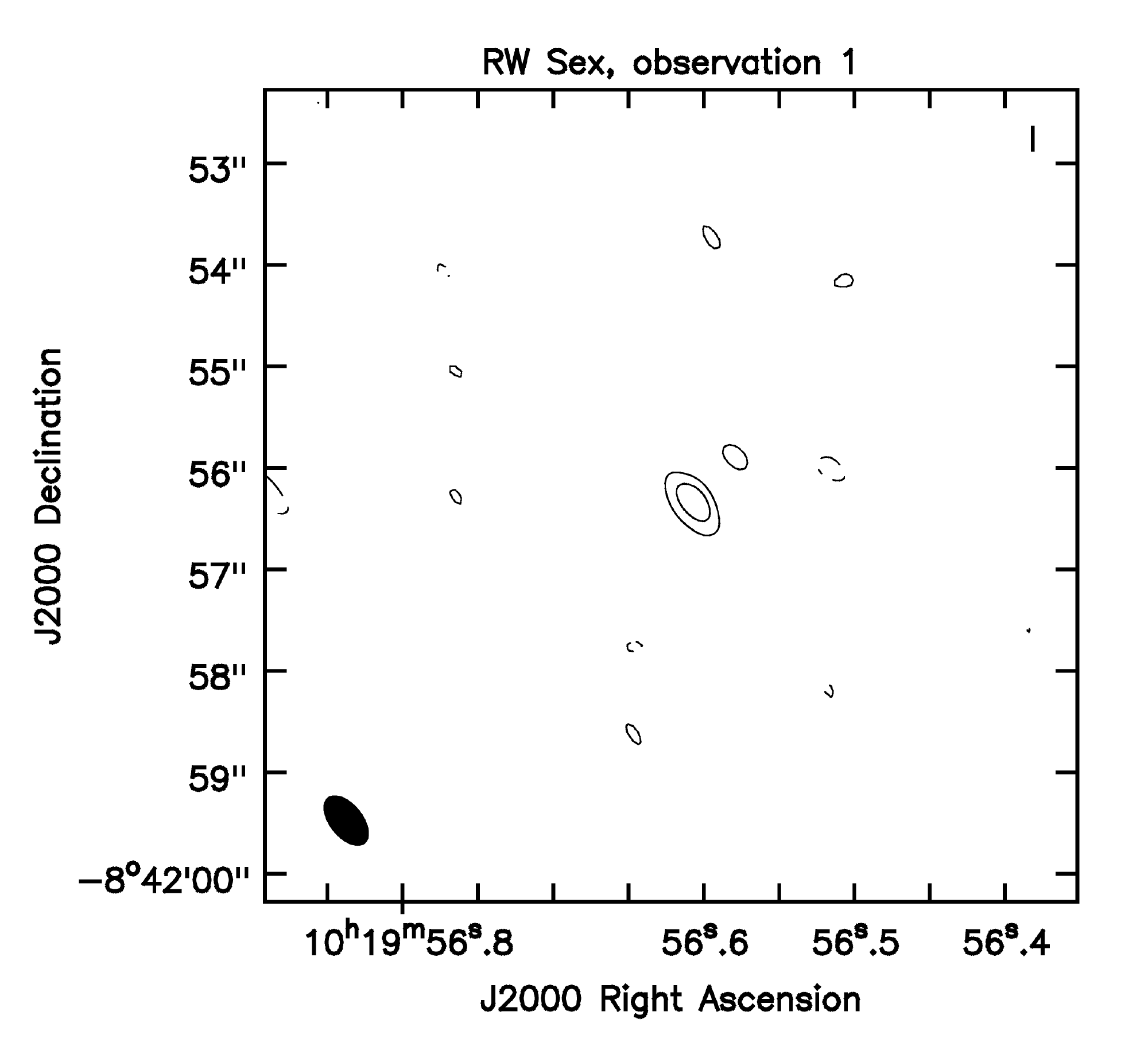}
    \label{fig:rwsex_obs1_I}
  \end{minipage}  
  \begin{minipage}{70mm}
    \centering
    \includegraphics[width=\textwidth]{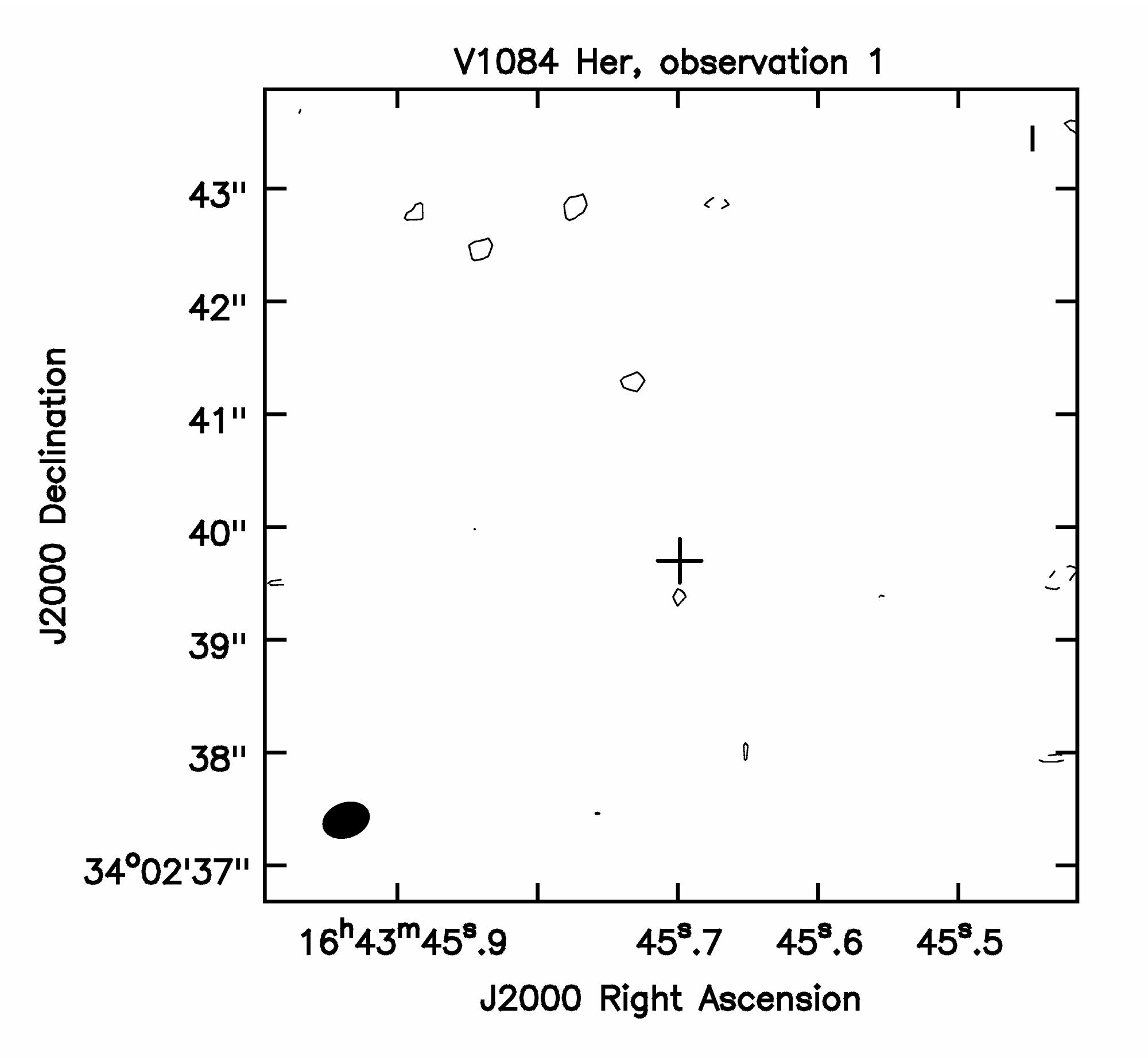}
    \label{fig:v1084her_obs1_I}
  \end{minipage}}\\[-2ex] 
  \subfloat{
  \begin{minipage}{70mm}
    \centering
    \includegraphics[width=\textwidth]{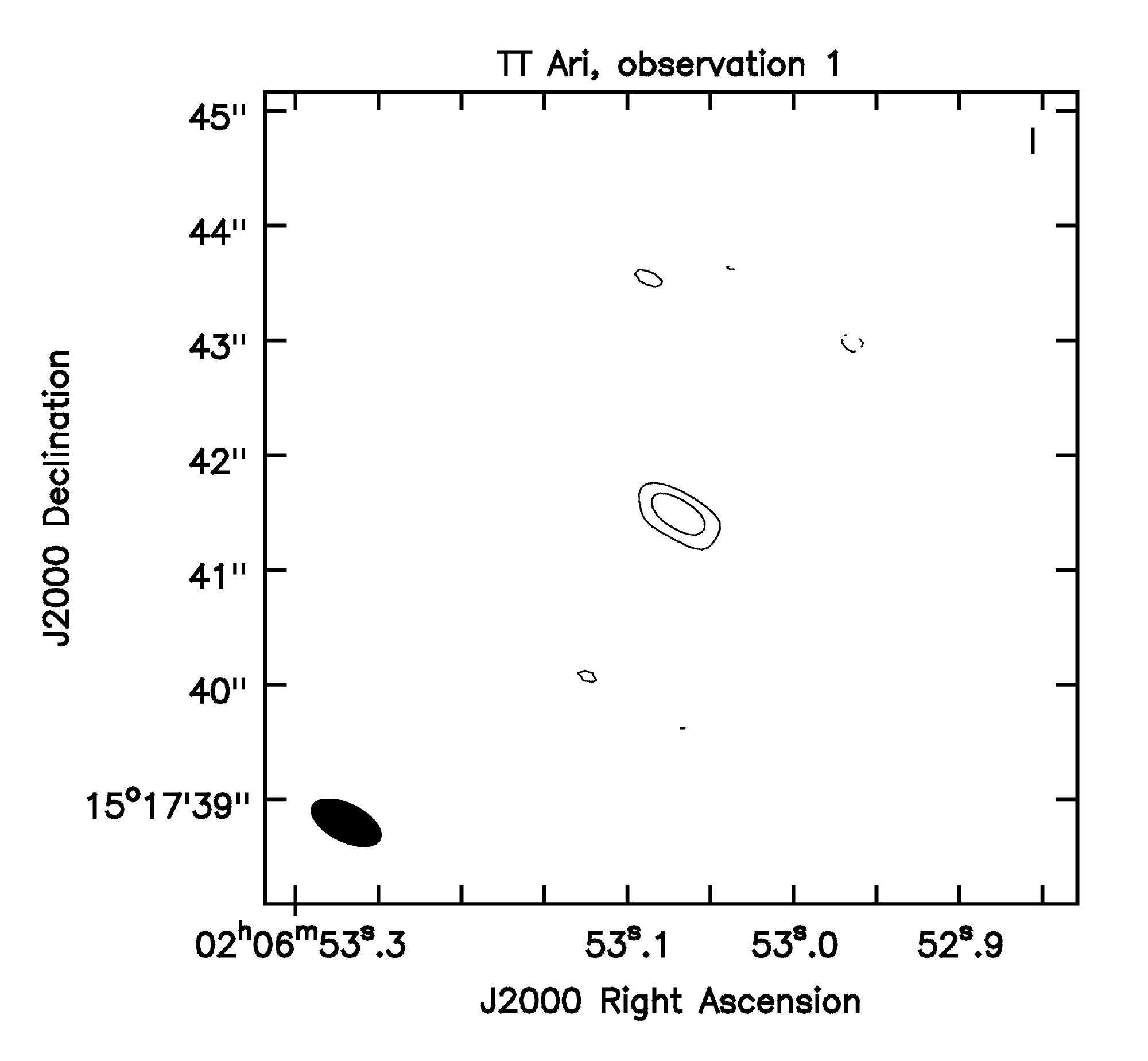}
    \label{fig:ttari_obs1_I}
  \end{minipage}  
  \begin{minipage}{70mm}
    \centering
    \includegraphics[width=\textwidth]{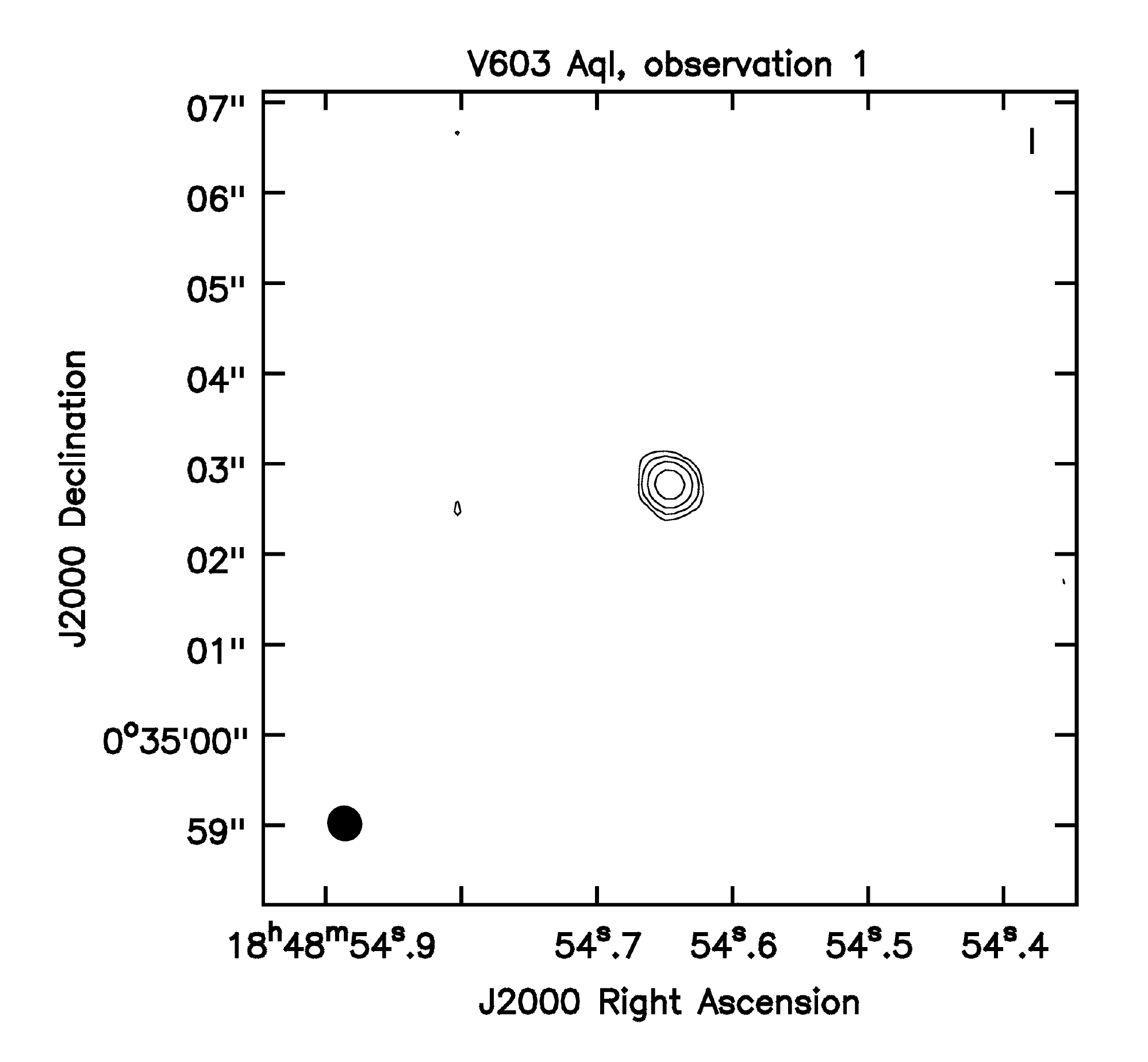}
    \label{fig:v603aql_obs1_I}
  \end{minipage}}\\[-2ex]
  \caption{Stokes I (total intensity) maps for the first observation of each novalike. Contours are at $\pm$3, $\pm$6, $\pm$12 and $\pm$24 sigma. The beam (resolution) is given in the lower left corner of each image. None of the detections show extended emission; they are all point sources. RMS and peak flux values are given in Table \ref{tbl:results}. The cross indicates the optical position for the non-detection (V1084 Her) -- the size is not indicative of the optical position error bars, as they are too small to be plotted here (see Table \ref{tbl:coords}).}
  \label{fig:stokesI}
\end{figure*}

\begin{figure*}
  \subfloat{
  \begin{minipage}{70mm}
    \centering
    \includegraphics[width=\textwidth]{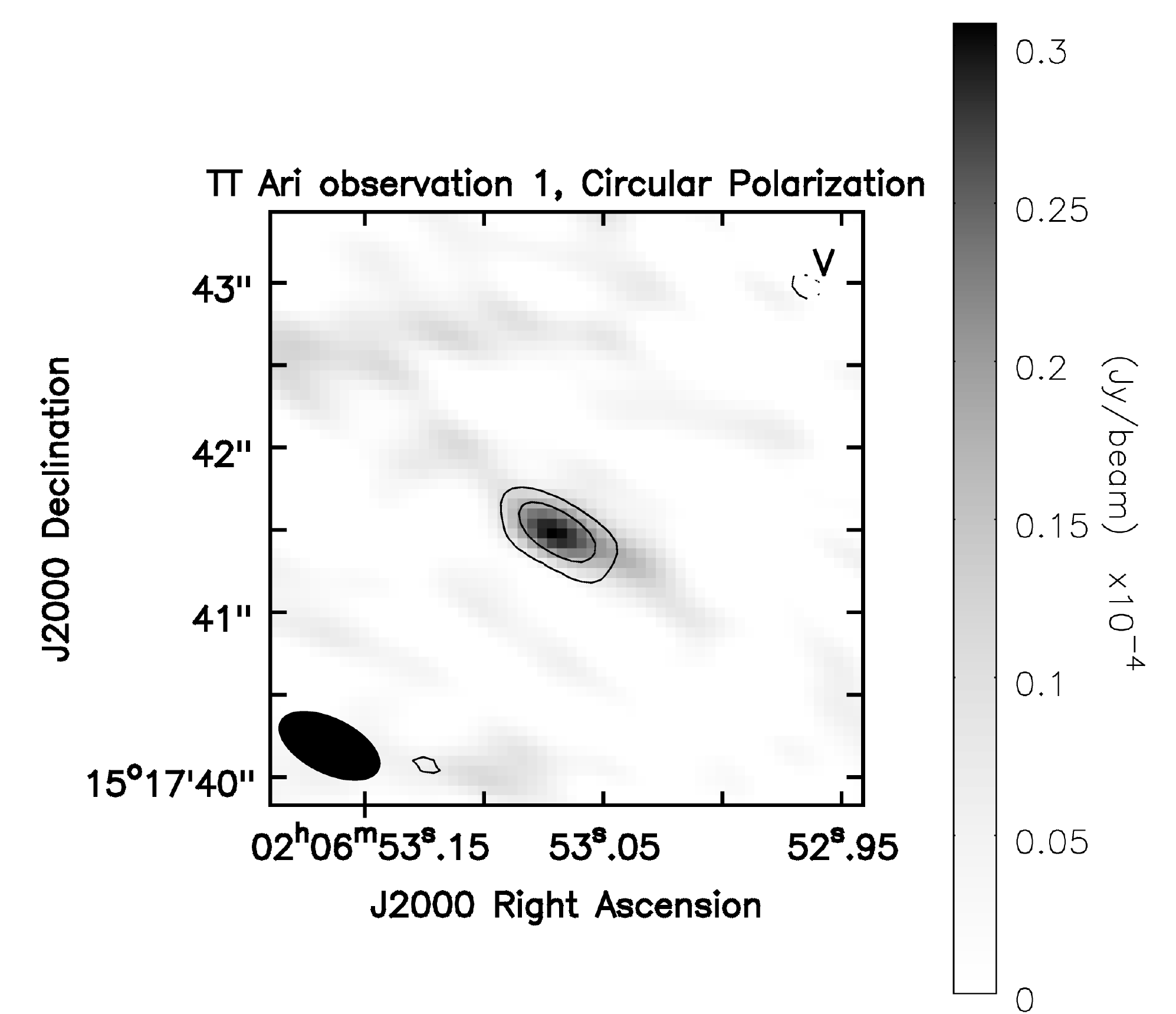}
  \end{minipage}  
  \begin{minipage}{70mm}
    \centering
    \includegraphics[width=\textwidth]{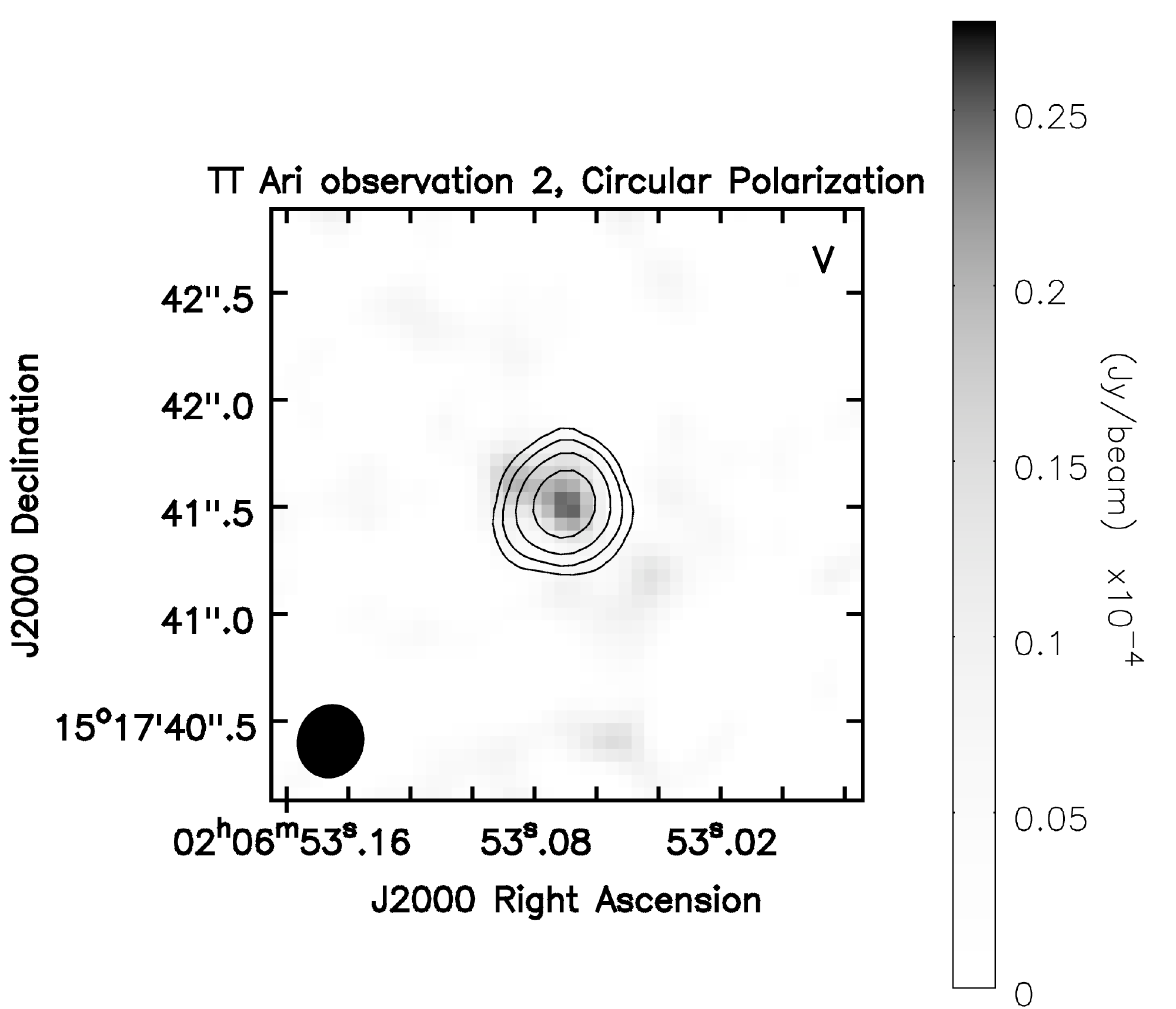}
    \label{fig:ttari_obs2_V}
  \end{minipage}}\\
  \caption{Stokes V (circular polarization) images for observation 1 (left) and 2 (right) of TT Ari. Stokes I contours are drawn at $\pm$3, $\pm$6, $\pm$12 and $\pm$24 sigma. For clarity purposes we do not show the left circularly polarized (negative) flux, as the detections were right circularly polarized (positive).}
  \label{fig:ttari_stokesV}
\end{figure*}

Here we show the total intensity and circular polarization maps (Figure \ref{fig:stokesI} and Figure \ref{fig:ttari_stokesV} respectively). See the captions for more information.

\end{document}